\documentclass[aps]{revtex4}

\usepackage{graphicx}

\begin{document}

\rightline{\today}

\title{Neutron Skin size dependence of the nuclear binding energy}

\author{S.J. Lee$^{1,2}$ and A.Z. Mekjian$^2$}

\affiliation
{$^1$Department of Applied Physics, Kyung Hee University, 
Yongin-si, Gyeonggi-do 446-701, Korea}

\affiliation
{$^2$Department of Physics and Astronomy, Rutgers University,
Piscataway, NJ 08854, USA}

\begin{abstract}
The nuclear binding energy is studied using a finite temperature density 
functional theory. A Skyrme interaction is used in this work. 
Volume, surface, and symmetry energy contributions to the binding 
energy are investigated.
The case of neutron skin is considered in detail.
The ratio of surface symmetry energy to volume symmetry energy of neutron 
skin dependent part is much larger than the corresponding ratio of neutron 
skin independent part. This shows that the large part of symmetry energy 
comes from the different size of neutron and proton distributions.
\end{abstract}



\maketitle


\section{Introduction}

The mass formulae \cite{ref1,ref7,dan9,dan11} characterizes the binding energy 
of nuclei in terms of the proton number $Z$ and neutron number $N$ or 
nucleon number $A=Z+N$. 
Volume, surface, Coulomb and pairing terms appear. Of importance and one of 
the least determined parts of the binding energy expression is the symmetry 
energy term. The symmetry energy has both a volume symmetry part and a 
surface symmetry part, 
similar to the division of the volume and surface terms. 
The division into volume and surface energies occurs because observed nuclei 
have finite nucleon number with a maximum value of $A \sim 250$. 
The symmetry energy appears in the total binding energy with factors involving 
both $((N-Z)/A)^2 A$ and $((N-Z)/A)^2 A^{2/3}$ for the volume and surface 
symmetry terms respectively. For a heavy nucleus with $A=200$, $Z=80$, $N=120$ 
the $(N-Z)/A=1/5$ and $((N-Z)/A)^2=1/25$. The isospin dependence of the binding 
energy arises in part from kinetic energy differences between protons and 
neutrons and from interaction terms arising from the isospin dependence of 
the nuclear force between nucleons. 
This interaction has terms involving the $\vec\tau_i\cdot\vec\tau_j$ which 
arises from the exchange of isovector mesons. The surface terms in the mass 
formulae arise from the loss of binding energy for nucleons near the surface 
of a nucleus. 
In a neutron star, the volume term only exists. Thus, of importance for neutron 
star physics is the extraction of the volume term from properties of known 
finite nuclei which contain both volume and surface terms. The symmetry term 
also plays a significant role in heavy ion collisions \cite{refa1,refa3,refa4},
in neutron star physics \cite{praka},
in the valley of nuclear stability and associated neutron 
and proton drip lines, 
in neutron halo nuclei in isobaric analog states \cite{refc,refd,refe},
and in giant isovector dipole states \cite{ref7}. 
Extensive discussion of the symmetry energy can be found in the work
of Danielewicz et al. \cite{daniel}. The role of the symmetry energy in 
the nuclear surface has also been studied in Ref.\cite{daniel}.
An extensive review can be found in Ref.\cite{praka}.

In our preveous paper \cite{prc82}, we have examined the $T$ dependence
of the expansion coefficients in mass formular by minimizing 
the Helmholtz free energy using Skyrme interaction.
In there we have assumed $R_n = R_p$ and thus missed any effects of
neutron skin which would exists in asymmetric nuclei.
We estimated approximately neutron skin effect for $T^2$ term in kinetic
energy by setting the central densities to be the same $\rho_{nc} = \rho_{pc}$
and found the effect of $R_n \ne R_p$ is small. 
But this is only a part of neutron skin effect thus we will study the 
neutron skin dependence more fully here by expanding energy
with the neutron skin size $R_n - R_p$.

In this paper we study the neutron skin effect on 
the volume and surface contributions to the symmetry 
energy. Our approach is based on a finite temperature density functional 
theory and we use a Skyrme type of interaction which we develop in the next 
section. The same Skyrme approach was also used to study phase 
transition in Ref.\cite{prc79,prc77,pl580,prc68}.
The symmetry energy and the neutron skin effects on symmetry energy and
other energy coefficients for three cases of Skyrme interactions are
studied in Sect. \ref{result} and 
the results are summarized in Sect. \ref{concl}.

\section{Binding energy in a density functional approach.}

A density functional theory based on a Skyrme interaction will be used in 
our investigation which is summarized in Appendix \ref{appena}. 
The density distribution is taken to be of the Saxon-Wood form:
\begin{eqnarray}
 \rho_q(\vec r) = \rho_q(R_q) = \frac{\rho_{qc}}{1 + e^{(r - R_q)/a_q}} 
                          \label{fermden}
\end{eqnarray}
The $q = p, n$ for protons, neutrons. 
We will allow the central density $\rho_{qc}$ and the $R_q$ for protons and 
neutrons to be different in general. 
The diffuseness $a_q$ are taken to be the same
for simplicity since our focus is on the neutron skin size 
dependence rather than the effects of different diffusness. 
Two limiting cases can also be considered. 
These are: 1. $R_p = R_n$ so that the difference in $N \ne Z$ nuclei is in 
the central density or $\rho_{pc} \ne \rho_{nc}$ 
and 2. $R_p \ne R_n$ and $\rho_{pc} = \rho_{nc}$. 
In case 2 the neutron distribution reflects a neutron halo for $N > Z$. 

Using this density functional approach the binding energy of nuclei as a 
function of mass number $A$, proton number $Z$ and temperature $T$ can be 
evaluated. The Weizacker semiemperical mass formulae \cite{ref1}, 
its extension by Myers and Swiatecki \cite{dan10,dan14,dan15,dan16} 
and also including finite temperature effects \cite{prc82} is 
\begin{eqnarray}
 E(A,Z,T) 
  &=& E_V(T) A + E_S(T) A^{2/3} + S_V(T) I^2 A + S_S(T) I^2 A^{2/3} 
            \nonumber \\    & &
      + E_C \frac{Z^2}{A^{1/3}} + E_{dif} \frac{Z^2}{A}
      + E_{ex} \frac{Z^{4/3}}{A^{1/3}} + c \Delta A^{-1/2}
                 \label{ldexpan}
\end{eqnarray}
where $I = (N-Z)/A = (A - 2Z)/A$.
The $B = -E_V$ is the usual bulk energy per nucleon.
The $E_{dif}$ and $E_{ex}$ are the coefficients for the diffuseness correction 
and the exchange correction to the Coulomb energy.  
For the pairing correction with constant $\Delta$, $c = +1$ for odd-odd
nuclei, 0 for odd-even nuclei, and $-1$ for even-even nuclei.
The above formula at $T =0$ is the well known Weizacker semiempirical mass 
formula \cite{ref1,ref7,dan9,dan11} studied extensively by Myers and 
Swiatecki \cite{dan10,dan14,dan15,dan16}. Early studies excluded the surface 
symmetry term $S_S$ and only the surface term $E_S$ was included. 
The values of the 
coefficients as found in textbooks such as Ref.\cite{ref7,dan11} 
are $E_V(0) = -B(0) \approx -16$, $E_S(0) \approx 17$, 
and $S_V(0) \approx 24$ in MeV. 
The ratio $E_S/B$ of surface to bulk energy at $T=0$ is very close to unity. 

Myers and Swiatecki \cite{dan10,dan14,dan15,dan16} have considered
an $A^{1/3}$ curvature term and a higher order $I^4$ term also.
However they dropped these two terms in their preliminary and illustrative
study of nuclear droplet model with arbitrary shape \cite{dan15,dan16}.
From the values in Ref.\cite{dan14} the curvature term is $7.0 A^{1/3}$ 
which is smaller than 1/6 of the surface term of $18.6 A^{2/3}$ for $A > 10$.
It should be noted that the so-called nuclear curvature energy 
puzzle \cite{npa563} concerns a higher theoretical value of the order 
of 10 MeV compared to a negligibly small empirical value for the
nuclear curvature energy.
On the other hand, the volume asymmetry energy is $28.1 I^2 A - 24.5 I^4 A$ 
\cite{dan14}. However typical values of the isospin asymmetry $I$ 
are smaller than 1/5.
Thus the term of $I^4$ is less than a few percent of $I^2$ term.
Furthermore for the energy of Eq.(\ref{hamilt}) an $I^4$ term comes 
directly only from kinetic energy. The simple kinetic energy can be
expanded as
\begin{eqnarray}
 N^{5/3} + Z^{5/3} \approx \left(\frac{A}{2}\right)^{5/3}
   \left(1 + \frac{5}{9} I^2 + \frac{5}{27} I^4 \right)    \nonumber 
\end{eqnarray}
Thus the ratio of the $I^4$ term compared to the $I^2$ term is $I^2/3$
which is about 1/75 for the lead region.
Since we are interested in the qualitative study of the energy expansion 
coefficients within the temperature and neutron skin size dependent part of 
the nuclear energy we did not include $A^{1/3}$ term and $I^4$ term here.

In our preveous paper \cite{prc82}, we have examined the $T$ dependence
of the expansion coefficients in Eq.(\ref{ldexpan}) by minimizing 
the Helmholtz free energy using a Skyrme interaction with the density
distribution of Eq.(\ref{fermden}).
With $R_n = R_p = R$, we can integrate Skyrme interaction analytically
to obtain total energy as a function of $R$. 
Using this energy function, the total energy $E(A, Z, T)$ minimizing 
free energy is found by varying the value of $R$ for a nucleus 
with $Z$ protons and $N$ neutrons at temprerature $T$.
Then we use Eq.(\ref{ldexpan}) for various nuclei to obtain the 
expansion coefficients.
Since we have assumed $R_n = R_p$ in our previous studies, 
we missed any effects of
neutron skin which would exist in asymmetric nuclei.
We estimated approximately neutron skin effect for $T^2$ term in kinetic
energy, Eq.(\ref{tauq}), 
by setting the central densities to be the same, 
 $\rho_{nc} = \rho_{pc} = \rho_c/2$ where $\rho_c$ is the total central
density obtained with $R_n = R_p$,
and found the effect of $R_n \ne R_p$ in kinetic energy is small.

In the present paper we now consider neutron skin effects more fully.
Specifically, we study the neutron 
skin size $t = R_n - R_p$ dependence of total energy $E(A, Z, T, t)$. 
Then we expand each coefficient $E_i$ of Eq.(\ref{ldexpan}) as
\begin{eqnarray}
 E_i(T, t) &=& E_i(T) + E_i(T)_{sk} \frac{|t|}{A^{1/3}}    \label{skexpan}
\end{eqnarray}
Notice here that $E_i(T,t)$, not $E_i(T)$, in Eq.(\ref{skexpan}) corresponds 
to $E_i(T)$ in Eq.(\ref{ldexpan}) which is the expansion coefficients
of empirical nuclear energy.
In Eq.(\ref{skexpan}), $E_i(T)$ (we use this notation in Eq.(\ref{skexpan}) 
just for simplicity) is the expansion coefficient of neutron skin
size $t$ independent part of nuclear energy (which is obtained by assuming
same neutron and proton distribution size $R_n = R_p = R$) 
and $E_i(T)_{sk}$ is the expansion coefficient of the first order $t$ 
dependent part of nuclear energy with $t/A^{1/3}$ factor.
Since we can integrate a Skyrme interaction analytically for the case
of $R_n = R_p = R$, we can obtain the neutron skin $t$ dependence
by expanding the integral for the case of $R_n \ne R_p$ 
around $R_n = R_p = R$.
This can be done by expanding the density distribution of Eq.(\ref{fermden})
around $R$ which we now discuss.
It should also be noted that in Eq.(\ref{skexpan}) the surface correction
appears with the factor $t/A^{1/3}$ which originates from
the dimensionless factor $t/R$ with $R = r_0 A^{1/3}$.

When $R_q \approx R$, the neutron skin size $t/R$ is proportional 
to $\frac{A^2}{NZ} = \frac{4}{(1 - I^2)}$ as can be seen 
in Appendix \ref{appenrho}.
Thus the skin dependence of the energy expansion coefficients have
this extra factor of $I$ dependence.
On the other hand, when neutron and proton central densities are the same, 
the neutron skin size $t/R$ is proportional to $(N-Z)/A$
(see Appendix \ref{appenrho}). 
Then the skin dependence of the energy expansion coefficients introduce 
an extra $I = (N-Z)/A$ behavior and
the energy expansion of Eq.(\ref{ldexpan}) with Eq.(\ref{skexpan})
becomes a third power expansion in the isospin factor $I$.
That is $t/A^{1/3} \sim I$ since $t/R \sim I$.
For such a case, by expanding the empirical nuclear energy for various nuclei 
together with $I$ and $I^3$ terms included, we may be able to extract the 
information about the neutron skin size from the odd term in $I$.
The odd term in $I$ coming from the skin size $t$ does not break isospin 
symmetry as we will discuss after Eq.(\ref{tqt}).
An odd power of $|I|$ was considered in Ref.\cite{oddi} also.

The central density $\rho_{qc}$ of the density distribution 
Eq.(\ref{fermden}) for a given value of $R_q$ should be determined 
to give a fixed number of nucleons $N_q$.
This normaization condition gives the expansion of $\rho_{qc}(R_q)$ 
as given by Eq.(\ref{rhoqc}) in Appendix \ref{appenrho} 
up to first order in $t_q = (R_q - R)$.
Thus the first order correction coefficient to $\rho_{qc}^m$ due to 
fixed $N_q$ is the zeroth order term times
 $-\frac{m}{R} \left[\frac{3 + \pi^2 \left(\frac{a}{R}\right)^2}
                   {1 + \pi^2 \left(\frac{a}{R}\right)^2}\right]$.
This result is independent of which type of particle, neutron or proton.
Furthermore if we keep the particle number $N_q$ and the total central 
density $\rho_c = \rho_{nc} + \rho_{pc}$ to be a constant 
while varying $R_q$, that is,
\begin{eqnarray}
 \rho_c(R) &=& \rho_{nc}(R_n) + \rho_{pc}(R_p)    
            = \rho_{nc}(R) + \rho_{pc}(R)
\end{eqnarray}
then we have 
\begin{eqnarray}
 t_n &=& R_n - R = \frac{Z}{A} t ,      \nonumber  \\
 t_p &=& R_p - R = - \frac{N}{A} t      \label{tqt}
\end{eqnarray}
to lowest order in $t = R_n - R_p = t_n - t_p$ when we expand 
about $R_q = R$ (see Appendix \ref{appenrho}).
For $N > Z$, $t > 0$ with $t_n > 0$ and $t_p < 0$.
For $Z > N$, $t < 0$ with $t_n < 0$ and $t_p > 0$.
Thus the roles of $t_n$ and $t_p$ are exchanged as $t$ changes sign.
Since the term $E_i(T)_{sk} t/ A^{1/3}$ without the absolute value 
sign in Eq.(\ref{skexpan}) comes from the expansion term 
 $\sum_q \left(\frac{d E(A,Z,T)}{dR_q}\right)_{R_q=R} t_q$ 
of the energy $E(A,Z,T)$,
exchanging the role of $t_n$ and $t_p$ does not break isospin symmetry of 
the nuclear energy. Due to the sign in Eq.(\ref{tqt}), $E_i(T)_{sk}$ changes 
sign as $t$ changes sign keeping the sign of the whole term unchanged.
However if we use $|t|$ with absolute value sign explicitely written 
instead of $t$ itself as used in Eq.(\ref{skexpan}) then $E_i(T)_{sk}$ 
does not change sign as $t$ changes sign. 
Since we considered here beta stable nuclei only with $N > Z$ 
we can drop the absolute value sign of $t$ in Eq.(\ref{skexpan}). 

The Fermi density $\rho_q(r)$ is then expanded about $R_q = R$ up to 
first order in $(R_q - R)$ as given by Eqs.(\ref{rhoqr})-(\ref{rhomnp}).
The results of Appendix \ref{appenrho} show that the
quantity $\rho_n^m(r) + \rho_p^m(r)$ has a first order correction 
from skin size $t$ and the first order correction vanishes for $m = 1$.
That is the total density $\rho(r) = \rho_n(r) + \rho_p(r)$ is independent
of the neutron skin size $t = R_n - R_p$ up to first order.
Thus only the explicit $\rho_q$ dependent terms in Skyrme interaction,
not the total density $\rho$ dependent terms, depend on the skin size $t$
up to the first order.

Since $F[\rho_q] = \int d^3r f(\rho_q(\vec r))$ for a Fermi density
 $\rho_q(r) = \frac{\rho_{qc}}{1 + e^{(r-R_q)/a}}$,
where $f(\rho_q)$ is only a function of single density $\rho_q$,
then $F[\rho_q]$ can be integrate exactly. 
The result is a function $F(R_q)$ of $R_q$ \cite{prc82}, 
which we can expand easily in terms of $t_q = R_q - R$ around $F(R)$.
That is
\begin{eqnarray}
 F(R_q) &=& F(R) + \frac{d F(R)}{d R} t_q
           + \frac{d^2 F(R)}{d R^2} \frac{t_q^2}{2} + \cdots
\end{eqnarray}
This procedure is much simpler than using a method of expanding the 
density in $t$ first and then integrating the results.
For fixed $N_q$, the central density $\rho_{qc}(R_q)$ is also a
function of $R_q$ as in Eq.(\ref{rhoqc}).
The expansion of integral of various power of $\rho_q$ are summarized
in Appendix \ref{appendc}.

For the integral of the form of $\int d^3 r \rho^\alpha \rho_q^m$,
we need to expand $\rho_q^m(R_q)$ around $R$ first then integrate
each term which is now a function of $R$ only. The results for various
cases are also given in Appendix \ref{appendc}.

\section{Netron skin size dependence}  \label{result}

Here we examine the neutron skin size dependence of nuclear energy 
using various Skyrme interactions for beta stable nuclei. 
We used three sets of Skyrme parameters with different values of the
effective mass in symmetric nuclear matter which are 
SLy4 with $m^*/m = 0.69$, SkM$^*$ with $m^*/m = 0.79$,
and SkM($m^*=m$) with $m^*/m = 1$.
The results for these three cases are given in Table \ref{tabl1}
and summarized in the following equations.
The three cases cover a wide range of interaction types in terms of
effective mass within the many various Skyrme interactions.

\begin{table}
\vspace*{-0.85in}
\caption{Neutron Skin Dependence of Energy coefficient
minimizing free energy. 
  }  \label{tabl1}
\hspace*{-0.99in}
\begin{tabular}{l|rrrr|rrrr|rrrr}
\hline
                  &   \multicolumn{4}{|c}{SLy4}           
                  &   \multicolumn{4}{|c}{SkM$^*$}        
                  &   \multicolumn{4}{|c}{SkM($m^*=m$)}     \\
\hline
 $T$ (MeV)       &  0 \ \ \ &  1 \ \ \ &  2 \ \ \ &  3 \ \ \ 
                 &  0 \ \ \ &  1 \ \ \ &  2 \ \ \ &  3 \ \ \  
                 &  0 \ \ \ &  1 \ \ \ &  2 \ \ \ &  3 \ \ \ \\
\hline
 $E_V(T)$        & --15.308 & --15.296 & --15.263 & --15.217 
                 & --15.127 & --15.108 & --15.051 & --14.962  
                 & --15.310 & --15.270 & --15.152 & --14.954 \\
 \ \ $T$-indp    & --15.308 & --15.308 & --15.308 & --15.312 
                 & --15.127 & --15.127 & --15.127 & --15.130 
                 & --15.310 & --15.310 & --15.310 & --15.310 \\
 \ \ $T^2$       &  0.01151 &  0.01145 &  0.01121 &  0.01061 
                 &  0.01925 &  0.01921 &  0.01908 &  0.01868 
                 &  0.03948 &  0.03948 &  0.03950 &  0.03948 \\
 \ \ Kine        &  28.653  &  28.594  &  28.423  &  28.124  
                 &  25.483  &  25.444  &  25.319  &  25.105  
                 &  19.957  &  19.978  &  20.037  &  20.134  \\
\ \ $E_V(T)_{sk}$&  5.5202  &  5.5100  &  5.4792  &  5.42745 
                 &  5.5837  &  5.5730  &  5.5411  &  5.48746  
                 &  6.9809  &  6.9701  &  6.9379  &  6.88390 \\
 \ \ \ \ $T$-indp&  5.5202  &  5.5172  &  5.5081  &  5.4927  
                 &  5.5837  &  5.5806  &  5.5717  &  5.5563  
                 &  6.9809  &  6.9780  &  6.9697  &  6.9553  \\
 \ \ \ \ $T^2$   &--0.00722 &--0.00722 &--0.00723 &--0.00725 
                 &--0.00761 &--0.00762 &--0.00763 &--0.00765  
                 &--0.00793 &--0.00793 &--0.00793 &--0.00794 \\
 \ \ \ \ Kine    & --3.8780 & --3.8654 & --3.8269 & --3.7576 
                 & --0.4893 & --0.4995 & --0.5302 & --0.5816  
                 & --0.3867 & --0.3963 & --0.4253 & --0.4737 \\
\hline
 $E_S(T)$        &  20.008  &  20.559  &  22.247  &  25.204  
                 &  18.756  &  19.296  &  20.948  &  23.836  
                 &  18.303  &  18.804  &  20.321  &  22.904  \\
 \ \ $T$-indp    &  20.008  &  20.009  &  20.024  &  20.099  
                 &  18.756  &  18.757  &  18.771  &  18.841  
                 &  18.303  &  18.304  &  18.310  &  18.339  \\
 \ \ $T^2$       &  0.5483  &  0.5501  &  0.5559  &  0.5672  
                 &  0.5371  &  0.5388  &  0.5443  &  0.5550  
                 &  0.4995  &  0.5003  &  0.5027  &  0.5072  \\
 \ \ Kine        & --32.106 & --31.638 & --30.255 & --27.943 
                 & --26.275 & --25.810 & --24.402 & --22.060  
                 & --15.845 & --15.379 & --13.966 & --11.598 \\
\ \ $E_S(T)_{sk}$& --17.077 & --17.040 & --16.928 & --16.739 
                 & --17.261 & --17.222 & --17.103 & --16.903  
                 & --21.691 & --21.656 & --21.550 & --21.372 \\
 \ \ \ \ $T$-indp& --17.077 & --17.059 & --17.006 & --16.915 
                 & --17.261 & --17.243 & --17.187 & --17.090  
                 & --21.691 & --21.678 & --21.638 & --21.569 \\
 \ \ \ \ $T^2$   &  0.01959 &  0.01958 &  0.01957 &  0.01955 
                 &  0.02082 &  0.02082 &  0.02082 &  0.02080  
                 &  0.02201 &  0.02201 &  0.02200 &  0.02199 \\
 \ \ \ \ Kine    &  12.325  &  12.290  &  12.179  &  11.981  
                 &  1.4093  &  1.4377  &  1.5238  &  1.6681  
                 &  1.1228  &  1.1493  &  1.2296  &  1.3638  \\
\hline
 $S_V(T)$        &  31.113  &  31.664  &  33.401  &  36.645  
                 &  29.655  &  30.210  &  31.960  &  35.221   
                 &  19.685  &  20.110  &  21.412  &  23.695  \\
 \ \ $T$-indp    &  31.113  &  31.115  &  31.150  &  31.333  
                 &  29.655  &  29.657  &  29.693  &  29.875  
                 &  19.685  &  19.686  &  19.698  &  19.756  \\
 \ \ $T^2$       &  0.5445  &  0.5489  &  0.5626  &  0.5902  
                 &  0.5482  &  0.5527  &  0.5667  &  0.5940  
                 &  0.4227  &  0.4241  &  0.4287  &  0.4377  \\
 \ \ Kine        & --24.674 & --24.568 & --24.103 & --23.455 
                 & --33.636 & --33.481 & --32.837 & --31.703   
                 & --6.8813 & --6.5197 & --5.4723 & --3.7475 \\
\ \ $S_V(T)_{sk}$&  63.203  &  63.225  &  63.266  &  63.306  
                 &  62.787  &  62.836  &  62.956  &  63.125  
                 &  63.366  &  63.381  &  63.428  &  63.498  \\
 \ \ \ \ $T$-indp&  63.203  &  63.194  &  63.140  &  63.019  
                 &  62.787  &  62.806  &  62.836  &  62.852  
                 &  63.366  &  63.350  &  63.301  &  63.209  \\
 \ \ \ \ $T^2$   &  0.03095 &  0.03104 &  0.03134 &  0.03188 
                 &  0.02954 &  0.02962 &  0.02988 &  0.03037  
                 &  0.03144 &  0.03151 &  0.03171 &  0.03207 \\
 \ \ \ \ Kine    &  1.3320  &  1.3578  &  1.3656  &  1.3511  
                 & --1.3400 & --1.2880 & --1.1223 & --0.8333  
                 & --0.8709 & --0.8224 & --0.6789 & --0.4350 \\
\hline
 $S_S(T)$        & --41.035 & --43.642 & --51.815 & --66.958 
                 & --43.596 & --46.185 & --54.328 & --69.369  
                 & --33.175 & --35.246 & --41.591 & --52.721 \\
 \ \ $T$-indp    & --41.035 & --41.046 & --41.196 & --41.984 
                 & --43.596 & --43.606 & --43.767 & --44.551  
                 & --33.175 & --33.179 & --33.234 & --33.522 \\
 \ \ $T^2$       & --2.5769 & --2.5962 & --2.6548 & --2.7749 
                 & --2.5598 & --2.5793 & --2.6404 & --2.7575  
                 & --2.0606 & --2.0668 & --2.0893 & --2.1332 \\
 \ \ Kine        &  171.020 &  170.079 &  165.833 &  159.697 
                 &  166.818 &  165.864 &  162.452 &  155.996  
                 &  56.5082 &  54.7216 &  49.7064 &  41.5131 \\
\ \ $S_S(T)_{sk}$& --711.559& --711.507& --711.128& --710.053
                 & --706.246& --706.252& --706.059& --705.169  
                 & --611.848& --612.332& --613.788& --616.101\\
 \ \ \ \ $T$-indp& --711.559& --710.590& --707.436& --701.662
                 & --706.246& --705.310& --702.271& --696.560 
                 & --611.848& --611.377& --609.957& --607.442\\
 \ \ \ \ $T^2$   & --0.91606& --0.91767& --0.92289& --0.93232
                 & --0.94030& --0.94192& --0.94708& --0.95653 
                 & --0.95456& --0.95538& --0.95785& --0.96219\\
 \ \ \ \ Kine    & --276.281& --275.271& --271.698& --265.302
                 & --40.3526& --41.5910& --45.3514& --51.7077  
                 & --31.5405& --32.7350& --36.3085& --42.3087\\
\hline
\end{tabular}
\end{table}

For Coulomb energy, from Eqs.(\ref{ecoul}) and (\ref{ecex}) 
with $a = 0.53$ fm and $R = 1.25 A^{1/3}$ fm,
$E_C$ part is 0.6912000, $E_{ex}$ part is --0.5278064 and $E_{diff}$ --1.430810
and the skin dependence of Coulomb energy 
$E_C$ part is --0.5529600, $E_{ex}$ part is 0.4222251 and $E_{diff}$ 3.433944,
that is
\begin{eqnarray}
 E_C(T,t) \frac{Z^2}{A^{1/3}} + E_{dif}(T,t) \frac{Z^2}{A}
      + E_{ex}(T,t) \frac{Z^{4/3}}{A^{1/3}}  
  &=& \left[0.6912 \frac{Z^2}{A^{1/3}} - 1.4308 \frac{Z^2}{A}
       - 0.5278 \frac{Z^{4/3}}{A^{1/3}}\right] 
              \nonumber  \\  & &     \hspace{-0.9in} 
    + \left[- 0.5530 \frac{Z^2}{A^{1/3}} + 3.4339 \frac{Z^2}{A}
      + 0.4222 \frac{Z^{4/3}}{A^{1/3}}\right] \frac{t}{A^{1/3}} 
\end{eqnarray}
in MeV and fm units.

In Table \ref{tabl1}, the items labeled ``$E_i(T)$'' and ``$E_i(T)_{sk}$''
are corresponding terms of Eq.(\ref{skexpan}) for
the energy expansion coefficients given in Eq.(\ref{ldexpan}). 
The items labeled by ``$T$-indp'', ``$T^2$'',
and ``Kine'' under $E_i(T)$ are the values of temperature $T$ independent
part, $T^2$ dependent term in kinetic energy (Eq.(\ref{tauq})),
and the total kinetic energy contribution to the energy expansion
coefficients $E_i(T)$ respectively.
Similarly the items labeled by ``$T$-indp'', ``$T^2$'', and ``Kine'' 
under $E_i(T)_{sk}$ are the values of temperature $T$ independent
part, $T^2$ dependent term in kinetic energy (Eq.(\ref{tauq})),
and the total kinetic energy contribution to the neutron skin dependent
part of the energy expansion coefficients $E_i(T)_{sk}$ respectively.

The results of Table \ref{tabl1} show the following features for
the skin size dependence of various coefficients $E_i(T)_{sk}$ 
of Eq.(\ref{skexpan}) in the mass formulae which behave 
as $E_i(T)_{sk} t/A^{1/3}$.
One overall feature for all components $E_i(T)_{sk}$ is the weak
dependence on temperature.
The volume binding energy coefficient $E_i(T)_{sk} = E_V(T)_{sk}$
of about $5 \sim 7$ MeV/fm has the smallest value of all the $E_i(T)_{sk}$
terms. Comparing to this value, the skin independent volume energy 
coefficient $E_V(T)$ is about $-15$ MeV.
(Notice here that the $E_i(T)_{sk} t/A^{1/3}$ with an unknown small value 
for the $t/A^{1/3}$ factor and not the $E_i(T)_{sk}$ itself should be 
compared to $E_i(T)$ for the value of energy coefficients.)
The surface energy coefficient $E_S(T)_{sk}$ has a somewhat larger values
of about $-17 \sim -22$ MeV/fm and is negative
compared to about 20 MeV for skin independent coefficient $E_S(T)$.
The neutron skin size dependence of the symmetry energy terms
have the following features.
The volume symmetry energy coefficient $S_V(T)_{sk} \sim 60$ MeV/fm 
while the surface symmetry energy coefficient has the largest magnitude
of about $-600 \sim - 700$ MeV/fm.
Comparing to these the neutron skin size independent coefficients
are $S_V(T) \approx 20 \sim 30$ MeV and $S_S(T) \approx 30 \sim 50$ MeV.
Some dependence on the effective mass is present for all coefficients
as can be seen in comparing SLy4 ($m^*/m=0.7$), SkM$^*$ ($m^*/m=0.8$),
and SkM($m^*=m$).
However the dependences of the individual coefficients on the 
temperatures and on the Skyrme parameters we used are not so sensitive.
The magnitude of the neutron skin size dependent coefficients 
are largely different between different coefficients
ranging from about 5 MeV/fm for volume energy coefficient
to about 700 MeV/fm for surface symmetry energy coefficient
while the magnitudes of the skin size independent parts were of the
same order of magnitude ranging from about 15 MeV for volume energy
to about 50 MeV for surface symmetry energy coefficient.
The neutron skin size dependence of surface symmetry energy is
much larger than the skin dependence of volume symmetry energy
with $A$ dependence included even for a large $A$ of over 200.
The ratio of the surface symmetry energy to the volume symmetry 
energy $S_S(T)_{sk}/S_V(T)_{sk} \sim 10$ for neutron skin size $t$ 
dependent part while $S_S(T)/S_V(T) \sim 1.5$ for neutron skin independent 
part of $R_n=R_p$. (Notice also that $E_S(T)_{sk}/E_V(T)_{sk} \sim 3$ compared 
to $E_S(T)/E_V(T) \sim 1.2$.) This show that the large part of symmetry energy 
comes from the different size of neutron and proton distributions. 
If we assume $t/A^{1/3} = 0.1$, then the ratio of total symmetry energy
 $[S_S(T) + S_S(T)_{sk} t/A^{1/3}]/[S_V(T) + S_V(T)_{sk} t/A^{1/3}] \sim 3$.
One dimensional semi infinite nuclear matter calculations show that 
this ratio is $S_S/S_V = 1 \sim 4$ for total symmetry energy \cite{praka}. 
We can also see that the neutron skin size dependent coefficients
for volume energy and surface energy $E_V(T)_{sk}$ and $E_S(T)_{sk}$ 
have opposite sign compare to the corresponding skin size independent 
coefficients $E_V(T)$ and $E_S(T)$,
while volume symmetry energy and surface symmetry energy $S_V(T)$ 
and $S_S(T)$ have the same sign for neutron skin size dependent 
coefficients and independent coefficients.

The kinetic energy contribution to the neutron skin size dependent 
and independent coefficients have a much 
more sensitive dependence on the Skyrme parameter set we used. 
The magnitude of kinetic energy contribution follows somewhat the
magnitude of the effective mass of the Skyrme parameter set used.
Since the neutron skin size dependent coefficients $E_i(T)_{sk}$
themselves are somewhat
insensitive to the Skyrme parameter set used, the potential energy
contribution to the coefficients, which are function of density,
also are sensitive to the parameter set used.
The kinetic energy contribution to the neutron skin size dependence 
$E_i(T)_{sk}$ are now opposite in sign to the kinetic energy part 
to the skin size independent coefficients $E_i(T)$ of energy 
expansion for all the terms. In turn, they are all opposite
sign to the skin size independent coefficients $E_i(T)$ themselves.
The kinetic energy contribution to the neutron skin size dependent
volume symmetry energy coefficient $S_V(T)_{sk}$ has a small magnitude 
similar to the magnitude of the kinetic energy contribution to the skin 
size dependent volume energy coefficient $E_V(T)_{sk}$.
The kinetic energy contribution to the surface symmetry energy coefficients, 
both the neutron skin size dependent and independent ones $S_S(T)_{sk}$
and $S_S(T)$, have largest
magnitude among the energy expansion coefficients. 
It is much larger even with $A$ dependence included than 
skin size dependence of volume symmetry energy $S_V(T)_{sk}$ 
and other energy expansion coefficients.

When the Helmholtz free energy is minimized, from the values 
of ``$T$-indp'' and ``$T^2$'' for $T = 0$ 
in Table \ref{tabl1}, the temperature and neutron skin size dependence
of the energy at low $T$ becomes
\begin{eqnarray}
 E(A, Z, T, t)  
  &=& \left[- (15.308 - 0.012 T^2) A + (20.008 + 0.548 T^2) A^{2/3} 
           \right.  \nonumber \\    & &   \hspace{0.3in}  \left. 
      + (31.113 + 0.545 T^2) I^2 A - (41.035 + 2.577 T^2) I^2 A^{2/3}\right] 
            \nonumber \\    & &
      + \left[(5.520 - 0.007 T^2) A - (17.077 - 0.020 T^2) A^{2/3} 
           \right.  \nonumber \\    & &   \hspace{0.3in}   \left.
      + (63.203 + 0.031 T^2) I^2 A - (711.559 + 0.916 T^2) I^2 A^{2/3} 
          \right] \frac{t}{A^{1/3}}
            \nonumber \\    & &
      + E_C(T,t) \frac{Z^2}{A^{1/3}} + E_{dif}(T,t) \frac{Z^2}{A}
      + E_{ex}(T,t) \frac{Z^{4/3}}{A^{1/3}}    
              \label{esly4}
\end{eqnarray}
for SLy4 parameter set,
\begin{eqnarray}
 E(A, Z, T, t)  
  &=& \left[- (15.127 - 0.019 T^2) A + (18.756 + 0.537 T^2) A^{2/3} 
           \right.  \nonumber \\    & &   \hspace{0.3in}  \left.
      + (29.655 + 0.548 T^2) I^2 A - (43.596 + 2.560 T^2) I^2 A^{2/3}\right] 
            \nonumber \\    & &
      + \left[(5.584 - 0.008 T^2) A - (17.261 - 0.021 T^2) A^{2/3} 
           \right.  \nonumber \\    & &   \hspace{0.3in}   \left.
      + (62.787 + 0.030 T^2) I^2 A - (706.246 + 0.940 T^2) I^2 A^{2/3} 
          \right] \frac{t}{A^{1/3}}
            \nonumber \\    & &
      + E_C(T,t) \frac{Z^2}{A^{1/3}} + E_{dif}(T,t) \frac{Z^2}{A}
      + E_{ex}(T,t) \frac{Z^{4/3}}{A^{1/3}}    
                  \label{eskms}
\end{eqnarray}
for SkM$^*$ parameter set, and
\begin{eqnarray}
 E(A, Z, T, t)  
  &=& \left[- (15.310 - 0.039 T^2) A + (18.303 + 0.500 T^2) A^{2/3} 
           \right.  \nonumber \\    & &   \hspace{0.3in}  \left.
      + (19.685 + 0.423 T^2) I^2 A - (33.175 + 2.061 T^2) I^2 A^{2/3}\right] 
            \nonumber \\    & &
      + \left[(6.981 - 0.008 T^2) A - (21.691 - 0.022 T^2) A^{2/3} 
           \right.  \nonumber \\    & &   \hspace{0.3in}   \left.
      + (63.366 + 0.031 T^2) I^2 A - (611.848 + 0.955 T^2) I^2 A^{2/3} 
          \right] \frac{t}{A^{1/3}}
            \nonumber \\    & &
      + E_C(T,t) \frac{Z^2}{A^{1/3}} + E_{dif}(T,t) \frac{Z^2}{A}
      + E_{ex}(T,t) \frac{Z^{4/3}}{A^{1/3}}    
                \label{eskmm}
\end{eqnarray}
for SkM($m^*=m$) parameter set in MeV and fm units.

Table \ref{tabl1} and above Eqs.(\ref{esly4})--(\ref{eskmm}) show that
the temperature $T$ dependent term of the energy expansion coefficients in 
the neutron skin dependent part have a simialr structure as the skin 
independent part. The volume energy $E_V(T)_{sk}$ has the smallest 
coefficent and $E_S(T)_{sk}$ and $S_V(T)_{sk}$ have similar magnitudes 
and the surface symmetry energy $S_S(T)_{sk}$ has the largest magnitude. 
The neutron skin size $t$ dependent part has a much smaller temperature $T$ 
dependence than the neutron skin size independent part.

From the values of ``$T$-indp'' and ``$T^2$'' for $T = 0$ 
in Table \ref{tabl1}, the temperature and 
neutron skin size dependence of the kinetic energy at low $T$ becomes
\begin{eqnarray}
 E_K(A, Z, T, t)  
  &=& \left[(28.653 + 0.012 T^2) A - (32.106 - 0.548 T^2) A^{2/3} 
           \right.  \nonumber \\    & &   \hspace{0.3in}  \left. 
      - (24.674 - 0.545 T^2) I^2 A + (171.020 - 2.577 T^2) I^2 A^{2/3}\right] 
            \nonumber \\    & &
      + \left[- (3.878 + 0.007 T^2) A + (12.325 + 0.020 T^2) A^{2/3} 
           \right.  \nonumber \\    & &   \hspace{0.3in}   \left.
      + (1.332 + 0.031 T^2) I^2 A - (276.281 + 0.916 T^2) I^2 A^{2/3} 
          \right] \frac{t}{A^{1/3}}           \label{ksly4}
\end{eqnarray}
for SLy4 parameter set,
\begin{eqnarray}
 E_K(A, Z, T, t)  
  &=& \left[(25.483 + 0.019 T^2) A - (26.275 - 0.537 T^2) A^{2/3} 
           \right.  \nonumber \\    & &   \hspace{0.3in}  \left.
      - (33.636 - 0.548 T^2) I^2 A + (166.818 - 2.560 T^2) I^2 A^{2/3}\right] 
            \nonumber \\    & &
      + \left[- (0.489 + 0.008 T^2) A + (1.409 + 0.021 T^2) A^{2/3} 
           \right.  \nonumber \\    & &   \hspace{0.3in}   \left.
      - (1.340 - 0.030 T^2) I^2 A - (40.353 + 0.940 T^2) I^2 A^{2/3} 
          \right] \frac{t}{A^{1/3}}           \label{kskms}
\end{eqnarray}
for SkM$^*$ parameter set, and
\begin{eqnarray}
 E_K(A, Z, T, t)  
  &=& \left[(19.957 + 0.039 T^2) A - (15.845 - 0.500 T^2) A^{2/3} 
           \right.  \nonumber \\    & &   \hspace{0.3in}  \left.
      - (6.881 - 0.423 T^2) I^2 A + (56.508 - 2.061 T^2) I^2 A^{2/3}\right] 
            \nonumber \\    & &
      + \left[- (0.387 + 0.008 T^2) A + (1.123 + 0.022 T^2) A^{2/3} 
           \right.  \nonumber \\    & &   \hspace{0.3in}   \left.
      - (0.871 - 0.031 T^2) I^2 A - (31.541 + 0.955 T^2) I^2 A^{2/3} 
          \right] \frac{t}{A^{1/3}}          \label{kskmm}
\end{eqnarray}
for SkM($m^*=m$) parameter set.

Table \ref{tabl1} and above Eqs.(\ref{esly4})--(\ref{kskmm}) show that the 
contribution of the $T^2$ dependent term in kinetic energy 
(see Eq.(\ref{tauq})) to the neutron skin size dependent 
coefficients $E_i(T)_{sk}$ are 
very small which is consistent with the result in Ref.\cite{prc82}.
In Ref.\cite{prc82}, we estimated approximately the neutron skin
size dependence coming from the $T^2$ term in kinetic energy
(see Eq.(\ref{tauq})) and found the effects were small.
The contribution of the $T^2$ dependent term in kinetic energy to the
neutron skin size are insensitive to the Skyrme parameter set used and
to the temperature.
The neutron skin dependent volume energy coefficient $E_V(T)_{sk}$ 
has smallest effect of $T^2$ dependence and the surface symmetry 
energy $S_S(T)_{sk}$ has a largest effect similar to the $T^2$ dependence 
of the corresponding skin independent coefficients of energy expansion.

By fitting the values of $E_i(T)_{sk}$ for temperatures which 
are $T=0$, 1, 2, and 3 MeV in Table \ref{tabl1}, the $T$-dependences 
of the neutron skin dependent coefficients are 
\begin{eqnarray}
 E_V(T)_{sk} &[=& 5.52016 - 0.00722 T^2]          \nonumber  \\
   &=& 5.52016 - 0.00008 T - 0.01005 T^2 - 0.00007 T^3  \label{evssly4}  \\
 E_S(T)_{sk} &[=& - 17.07720 + 0.01959 T^2]   \nonumber  \\
   &=& - 17.07720 + 0.00091 T + 0.03599 T^2 + 0.00044 T^3 \label{esssly4} \\
 S_V(T)_{sk} &[=& 63.20304 + 0.03095 T^2]   \nonumber  \\
   &=& 63.20304 + 0.00560 T + 0.01927 T^2 - 0.00322 T^3  \label{svssly4} \\
 S_S(T)_{sk} &[=& - 711.5588 - 0.91606 T^2]   \nonumber \\
   &=& - 711.5588 + 0.00963 T - 0.01940 T^2 + 0.06117 T^3  \label{ssssly4}
\end{eqnarray}
for SLy4 parameter set.
The first expressions given in the square parenthesis are from the values 
for $T=0$ in Table \ref{tabl1} same as in Eqs.(\ref{esly4})-(\ref{kskmm}) 
for comparison.
\begin{eqnarray}
 E_V(T)_{sk} &[=& 5.58368 - 0.00761 T^2]      \nonumber  \\
   &=& 5.58368 - 0.00042 T - 0.01018 T^2 - 0.00013 T^3   \label{evsskms} \\
 E_S(T)_{sk} &[=& - 17.26147 + 0.02082 T^2]   \nonumber  \\
   &=& - 17.26147 + 0.00154 T + 0.03763 T^2 + 0.00056 T^3 \label{essskms} \\
 S_V(T)_{sk} &[=& 62.78658 + 0.02954 T^2]   \nonumber  \\
   &=& 62.78658 + 0.00634 T + 0.04623 T^2 - 0.00358 T^3  \label{svsskms}  \\
 S_S(T)_{sk} &[=& - 706.2461 - 0.94030 T^2]   \nonumber \\
   &=& - 706.2461 + 0.06100 T - 0.15020 T^2 + 0.08320 T^3  \label{sssskms}
\end{eqnarray}
for SkM$^*$ parameter set.
\begin{eqnarray}
 E_V(T)_{sk} &[=& 6.98087 - 0.00793 T^2]      \nonumber  \\
   &=& 6.98087 - 0.000120 T - 0.01048 T^2 - 0.00008 T^3  \label{evsskmm} \\
 E_S(T)_{sk} &[=& - 21.69093 + 0.022014 T^2]   \nonumber  \\
   &=& - 21.69093 + 0.00070 T + 0.03447 T^2 + 0.00026 T^3  \label{essskmm} \\
 S_V(T)_{sk} &[=& 63.36637 + 0.03144 T^2]   \nonumber  \\
   &=& 63.36637 - 0.00421 T + 0.02050 T^2 - 0.00150667 T^3  \label{svsskmm} \\
 S_S(T)_{sk} &[=& - 611.8476 - 0.95456 T^2]   \nonumber \\
   &=& - 611.8476 + 0.03905 T - 0.54265 T^2 + 0.01900 T^3   \label{sssskmm}
\end{eqnarray}
for SkM($m^*=m$) parameter set.

By fitting the values for $T=0$, 1, 2, and 3 MeV in Table \ref{tabl1},
the temperature $T$ and neutron skin size $t$ dependence of the expansion 
coefficients of the kinetic energy are
\begin{eqnarray}
 E_V(T, t)_K &[=& (28.65250 + 0.01151 T^2)
             + (- 3.87804 - 0.00722 T^2) \frac{t}{A^{1/3}} ]   \nonumber  \\
   &=& (28.65250 - 0.00801 T - 0.04819 T^2 - 0.00264 T^3)
                \nonumber  \\  & &
     + (- 3.87804 + 0.00117 T + 0.01064 T^2 + 0.00078 T^3)
             \frac{t}{A^{1/3}}        \label{evksly4}  \\
 E_S(T, t)_K &[=& (- 32.10626 + 0.54833 T^2)
             + (12.32541 + 0.01959 T^2) \frac{t}{A^{1/3}} ]   \nonumber  \\
   &=& (- 32.10626 + 0.01492 T + 0.45071 T^2 + 0.00230 T^3)
                \nonumber  \\  & &
     + (12.32541 - 0.00188 T - 0.03149 T^2 - 0.00207 T^3)
             \frac{t}{A^{1/3}}        \label{esksly4}  \\
 S_V(T, t)_K &[=& (- 24.67405 + 0.54454 T^2) 
             + (1.33205 + 0.03095 T^2) \frac{t}{A^{1/3}} ]   \nonumber  \\
   &=& (- 24.67405 - 0.13177 T + 0.26686 T^2 - 0.02916 T^3)
                \nonumber  \\  & &
     + (1.33205 + 0.03334 T - 0.00683 T^2 - 0.00072 T^3)
             \frac{t}{A^{1/3}}        \label{svksly4}  \\
 S_S(T, t)_K &[=& (171.0202 - 2.57693 T^2)
             + (- 276.2811 - 0.91606 T^2) \frac{t}{A^{1/3}} ]   \nonumber  \\
   &=& (171.0202 + 1.18275 T - 2.36020 T^2 + 0.23595 T^3)
                \nonumber  \\  & &
     + (- 276.2811 - 0.18403 T + 1.15055 T^2 + 0.04358 T^3)
             \frac{t}{A^{1/3}}            \label{ssksly4}
\end{eqnarray}
for SLy4 parameter set.
\begin{eqnarray}
 E_V(T, t)_K &[=& (25.48323 + 0.01925 T^2)
             + (- 0.48934 - 0.00761 T^2) \frac{t}{A^{1/3}} ]   \nonumber  \\
   &=& (25.48323 + 0.00253 T - 0.04169 T^2 - 0.00038 T^3)
                \nonumber  \\  & &
     + (- 0.48934 + 0.00007 T - 0.01019 T^2 - 0.00003 T^3)
             \frac{t}{A^{1/3}}       \label{evkskms}  \\
 E_S(T, t)_K &[=& (- 26.27463 + 0.53708 T^2)
             + (1.40926 + 0.02082 T^2) \frac{t}{A^{1/3}} ]   \nonumber  \\
   &=& (- 26.27453 - 0.00990 T + 0.47594 T^2 - 0.00145 T^3)
                \nonumber  \\  & &
     + (1.40926 - 0.00022 T + 0.02858 T^2 + 0.00008 T^3)
             \frac{t}{A^{1/3}}       \label{eskskms}  \\
 S_V(T, t)_K &[=& (- 33.63624 + 0.54816 T^2)
             + (- 1.33997 + 0.02954 T^2) \frac{t}{A^{1/3}} ]   \nonumber  \\
   &=& (- 33.63624 - 0.08819 T + 0.24295 T^2 + 0.00042 T^3)
                \nonumber  \\  & &
     + (- 1.33997 - 0.00176 T + 0.05211 T^2 + 0.00159 T^3)
             \frac{t}{A^{1/3}}       \label{svkskms}  \\
 S_S(T, t)_K &[=& (166.8184 - 2.55977 T^2) 
             + (- 40.35256 - 0.94030 T^2) \frac{t}{A^{1/3}} ]    \nonumber  \\
   &=& (166.8184 + 0.07982 T - 0.93650 T^2 - 0.09752 T^3) 
                \nonumber  \\  & &
     + (- 40.35256 - 0.00212 T - 1.22403 T^2 - 0.01232 T^3)
             \frac{t}{A^{1/3}}           \label{sskskms}
\end{eqnarray}
for SkM$^*$ parameter set.
\begin{eqnarray}
 E_V(T, t)_K &[=& (19.95733 + 0.03948 T^2) 
             + (- 0.38673 - 0.00793 T^2) \frac{t}{A^{1/3}} ]     \nonumber  \\
   &=& (19.95733 + 0.00161 T + 0.01917 T^2 - 0.00003 T^3) 
                \nonumber  \\  & &
     + (- 0.38673 + 0.00006 T - 0.00962 T^2 - 0.00002 T^3)
             \frac{t}{A^{1/3}}         \label{evkskmm}  \\
 E_S(T, t)_K &[=& (- 15.84463 + 0.49952 T^2) 
             + (1.12277 + 0.02201 T^2) \frac{t}{A^{1/3}} ]     \nonumber  \\
   &=& (- 15.84463 - 0.00388 T + 0.46826 T^2 + 0.00162 T^3) 
                \nonumber  \\  & &
     + (1.12277 - 0.00024 T + 0.02672 T^2 + 0.00005 T^3)
             \frac{t}{A^{1/3}}         \label{eskskmm}  \\
 S_V(T, t)_K &[=& (- 6.88128 + 0.42271 T^2) 
             + (- 0.87085 + 0.03144 T^2) \frac{t}{A^{1/3}} ]   \nonumber  \\
   &=& (- 6.88128 + 0.01595 T + 0.34709 T^2 - 0.00140 T^3) 
                \nonumber  \\  & &
     + (- 0.87085 + 0.00273 T + 0.04486 T^2 + 0.00089 T^3)
             \frac{t}{A^{1/3}}         \label{svkskmm}  \\
 S_S(T, t)_K &[=& (56.50820 - 2.06058 T^2) 
             + (- 31.54054 - 0.95456 T^2) \frac{t}{A^{1/3}} ]    \nonumber  \\
   &=& (56.50820 - 0.15544 T - 1.63960 T^2 + 0.00843 T^3) 
                \nonumber  \\  & &
     + (- 31.54054 - 0.02086 T - 1.16564 T^2 - 0.00796 T^3)
             \frac{t}{A^{1/3}}           \label{sskskmm}
\end{eqnarray}
for SkM($m^*=m$) parameter set.

Comparing the first line (or Eqs.(\ref{esly4})-(\ref{kskmm})) and 
second line of above Eqs.(\ref{evssly4}) -- (\ref{sskskmm}),
we can see the temperature $T$ dependence of the neutron skin size 
dependent coefficients $E_i(T)_{sk}$ comes not only from the $T^2$ term in
the kinetic energy, Eq.(\ref{tauq}), but also from the potential energy 
through the different saturation density for different temperature.
Especially, the neutron skin size dependence of kinetic energy expansion
coefficients for SLy4 parameter set have opposite sign in their $T$ 
dependence compared with the $T^2$ term in kinetic energy ($T^2$ 
dependence in Eqs.(\ref{esly4})-(\ref{kskmm})).
Table \ref{tabl1} and Eqs.(\ref{evssly4})-(\ref{sssskmm}) show that 
the $T$ dependence of the neutron skin size dependent coefficient is 
much faster than the $T^2$ term in kinetic energy of Eq.(\ref{tauq}) 
alone for volume energy coefficient $E_V(T)_{sk}$ 
and surface energy coefficient $E_S(T)_{sk}$.
By contrast this it is much slower than the $T^2$ term for volume 
symmetry energy coefficient $S_V(T)_{sk}$ and surface symmetry energy 
coefficient $S_S(T)_{sk}$ except for volume symmetry energy $S_V(T)_{sk}$ 
of the SkM$^*$ parameter set.

Eqs.(\ref{evksly4})-(\ref{sskskmm}) show that even the kinetic energy
also has an extra $T$ dependence, 
beside the $T^2$ term in kinetic energy of Eq.(\ref{tauq}),
through the $T$ dependence of saturation density.
For SLy4 Skyrme interaction, the $T$ dependence of the kinetic energy part
of the neutron skin size dependent energy expansion coefficients $E_i(T)_{sk}$
has opposite sign with the $T^2$ term of kinetic energy Eq.(\ref{tauq}).
For this interaction the $T$ dependence of the kinetic energy part
of the volume symmetry energy coefficient $S_V(T)_{sk}$ is rather linear
as compared to a $T^2$ behavior.
For SkM$^*$ and SkM($m^*=m$) interactions, the kinetic energy part of
the neutron skin dependent energy coefficients $E_i(T)_{sk}$ has the same 
sign in its $T$ dependences with the $T^2$ term in the kinetic energy 
of Eq.(\ref{tauq}) but has a faster dependence of $T$ compared to 
the $T^2$ term of Eq.(\ref{tauq}).
The neutron skin independent kinetic energy expansion coefficients have a
slower $T$ dependence than a $T^2$ term for a kinetic energy, Eq.(\ref{tauq}),
except the volume energy coefficient $E_V(T, t=0)_K$ for SLy4 and SkM$^*$
parameter sets which have an opposite sign compared to $T^2$ term of 
kinetic energy Eq.(\ref{tauq}).
The neutron skin size independent volume symmetry energy coefficient
 $S_V(T, t=0)_K$ for SLy4 parameter set has a linear $T$ dependence
comparable order to the $T^2$ dependence.

The kinetic energy expansion at zero $T$, from Eqs.(\ref{ksly4})-(\ref{kskmm})
or from Eqs.(\ref{evksly4})-(\ref{sskskmm}), are
\begin{eqnarray}
 E_K(A, Z, T=0, t)  
  &=& \left[28.653 A - 32.106 A^{2/3} 
          - 24.674 I^2 A + 171.020 I^2 A^{2/3}\right] 
            \nonumber \\    & &
      + \left[- 3.878  A + 12.325 A^{2/3} 
              + 1.332 I^2 A - 276.281 I^2 A^{2/3}\right] \frac{t}{A^{1/3}} 
                    \label{ksly4t0}
\end{eqnarray}
for SLy4 parameter set,
\begin{eqnarray}
 E_K(A, Z, T=0, t)  
  &=& \left[25.483 A - 26.275 A^{2/3} 
          - 33.636 I^2 A + 166.818 I^2 A^{2/3}\right] 
            \nonumber \\    & &
      + \left[- 0.489 A + 1.409 A^{2/3} 
              - 1.340 I^2 A - 40.353 I^2 A^{2/3}\right] \frac{t}{A^{1/3}} 
                     \label{kskmst0}
\end{eqnarray}
for SkM$^*$ parameter set, and
\begin{eqnarray}
 E_K(A, Z, T=0, t)  
  &=& \left[19.957 A - 15.845 A^{2/3} 
          - 6.881 I^2 A + 56.508 I^2 A^{2/3}\right] 
            \nonumber \\    & &
      + \left[- 0.387 A + 1.123 A^{2/3} 
              - 0.871 I^2 A - 31.541 I^2 A^{2/3}\right] \frac{t}{A^{1/3}} 
                     \label{kskmmt0}
\end{eqnarray}
for SkM($m^*=m$) parameter set. For an infinite nuclear matter, 
the surface terms disappear and only the volume terms survive.
These results show that the surface symmetry energy coefficient has the 
largest effect from kinetic energy as compared to the other coefficients.
Here we can see the surface kinetic energy and the volume symmetry kinetic 
energy coefficients are negative.
However the total kinetic energy and total symmetry kinetic energy
including $A$ factors are positive.
The neutron skin dependent kinetic energies with $A$ factor included 
are negative.
Compre to result from Fermi gas model,
\begin{eqnarray}
 E_K = 12 I^2 A + 9 I^2 A^{2/3}
\end{eqnarray}
which is good for high $T$ or low density limit without any interaction.
Here both the volume and surface symmetry energies are positive.
With Skyrme interaction, the isospin dependent part of the effective
mass in a finite nuclei may become negative depending on the force
parameter and densities of proton and neutron. 
Thus the signs in Eqs.(\ref{ksly4t0})-(\ref{kskmmt0}) result.

\section{Conclusion and summary}  \label{concl}

Understanding properties of the symmetry energy is important in many area 
of nuclear physics as mentioned in the introduction. 
In this paper we studied properties of the symmetry energy, 
both volume and surface parts, along with other terms which appear 
in the Weizsacker mass formulae. 
Our investigation was based on a finite temperature density functional approach.
In a finite temperature approach, the dependence of various terms on 
temperature can be obtained. Energetic probes lead to excited nuclei 
which may be characterized by a hot liquid drop extension of the Weizsacker 
mass formulae. We used several different interactions of the Skyrme type 
to examine the dependence of various quantities on the interaction and 
associated effective masses that appear. 
Our analysis in the present study emphasized the role of the neutron skin 
on various terms that appear in the mass formulae. 
The radii of protons and neutrons were therefore allowed to be different 
in a Saxon-Wood form for the density distributions of these particles. 
We then proceeded to calculate various terms using an expansion about the 
equal radii point. The corrections that arise from a neutron skin are 
then proportional to the skin thickness $t$ over the radius $R$ 
or $t/R \sim t/A^{1/3}$. The skin thickness $t/R$ itself can be proportional 
to the neutron excess fraction $I = (N-Z)/A$ when the proton and neutron
central densities are the same. 
On the other hand, if $R_n = R_p = R$ then $t/R$ is proportional 
to $A^2/ZN = 4/(1 - I^2)$.
Thus the $I$ dependences of various terms in the mass formulae are modified. 

Table \ref{tabl1} contains the results for three Skyrme interactions, 
SLy4, SkM$^*$ and SkM($m^*=m$).
Results for the volume energy $E_V(T)$ and $E_V(T)_{sk}$, 
surface energy $E_S(T)$ and $E_S(T)_{sk}$, 
volume symmetry energy $S_V(T)$ and $S_V(T)_{sk}$, 
and surface symmetry energy $S_S(T)$ and $S_S(T)_{sk}$ are given. 
The terms with an additional subscript ``$sk$'' are the skin coefficients 
of Eq.(\ref{skexpan}) and the terms without subscript ``$sk$'' are
the skin independent part ($t=0$) in Eq.(\ref{skexpan}). 
The kinetic energy contributions, labeled ``Kine'', 
to each term are also given. The difference of the total and kinetic term 
is from the interaction.

The results show that the neutron skin size dependent and independent 
energy expansion coefficients are rather insensitive to the Skryme 
interaction used while the kinetic energy and potential
energy expansion coefficients separately are sensitive to the interaction
used and somewhat follow the effective mass of the Skyrme parameter.
The temperature dependence of the neutron skin size dependent energy 
expansion coefficients are much more insensitive than the temperature
dependence of neutron skin size independent coefficients.
The magnitude of the neutron skin size dependent coefficients are
largely different for different coefficients compared to the
neutron skin size independent coefficients.
The neutron skin size dependent volume energy coefficient has the 
smallest magnitute while the surface symmetry energy coefficient has 
the largest magnitude.
The neutron skin size dependence of surface symmetry energy is much
larger than the skin size dependence of the volume symmetry energy
with $A$ dependence included.
The ratio of the surface symmetry energy coefficient to the volume 
symmetry energy coefficient is $S_S(T)_{sk}/S_V(T)_{sk} \sim 10$
for neutron skin dependent part compared to $S_S(T)/S_V(T) \sim 1.5$
for neutron skin independent part. 
This shows that the large part of symmetry energy comes from the 
different size of proton and neutron distributions.
The surface symmetry kinetic energy coefficients,
both the neutron skin size dependent and independent ones, have
the largest magnitude of the kinetic energy expansion coefficients.
The neutron skin size dependent volume energy coefficient has the smallest
temperatutre dependence and the neutron skin size dependent surface symmetry
energy coefficient has the largest temperature dependence similar 
to the neutron skin independent coefficients. 
The temperature dependence of the neutron skin size dependent coefficients
is smaller than the temparature dependence of the skin size independent
coefficients.
The temperature dependences of the neutron skin size dependent energy
coefficients have a large effect from the different saturation
density for different temperature and thus do not follow $T^2$ 
behavior of the explicit $T^2$ dependence of kinetic energy.

Considering the neutron skin size dependence $t/R$ factor 
with $R = r_0 A^{1/3}$, the neutron skin dependent energy expansion
coefficients has a factor of $t/A^{1/3}$ in the energy expansion.
If we relate the dimensionless factor $t/R$
to the isospin factor $I = (N-Z)/A$ then the energy expansion
including the neutron skin size dependence introduce an extra $I$
factor. With this extra $I$ dependence we may be able to extract some 
information on the neutron skin size by expanding the empirical energy of 
various nuclei with including odd power of $I$ up to third order
if the neutron and proton central densities are the same.

This work was supported in part 
by Grant No. KHU-20080646 of the Kyung Hee University Research Fund in 2008 
and by the US Department of Energy under DOE Grant No. DE-FG02ER-40987.

\appendix

\section{Skyrme interaction}\label{appena}

The Hamiltonian for a Skyrme interaction is 
\begin{eqnarray}
 H(\vec r) &=& H_B(\vec r) + H_S(\vec r) + H_C(\vec r)  \nonumber  \\
 H_B &=& \frac{\hbar^2}{2 m_p} \tau_p + \frac{\hbar^2}{2 m_n} \tau_n
                 \nonumber  \\
  & & + \frac{1}{4} \left[ t_1 \left(1 + \frac{x_1}{2}\right) 
           + t_2 \left(1 + \frac{x_2}{2}\right) \right] \rho \tau 
      - \frac{1}{4}\left[ t_1 \left(\frac{1}{2} + x_1\right)
           - t_2 \left(\frac{1}{2} + x_2\right) \right]  
         \left(\rho_p \tau_p + \rho_n \tau_n\right)
                 \nonumber  \\
  & & + \frac{t_0}{2} \left[ \left(1 + \frac{x_0}{2}\right) \rho^2 
    - \left(\frac{1}{2} + x_0\right) \left(\rho_p^2 + \rho_n^2\right) \right] 
                 \nonumber  \\
  & & + \frac{t_3}{12} \left[ \left(1 + \frac{x_3}{2}\right) \rho^2 
         - \left(\frac{1}{2} + x_3\right) \left(\rho_p^2 + \rho_n^2\right) 
        \right] \rho^{\alpha}
                 \nonumber  \\
 H_S &=& \frac{1}{16} \left[ 3 t_1 \left(1 + \frac{x_1}{2}\right)
                   - t_2 \left(1 + \frac{x_2}{2}\right) \right]
                (\vec\nabla\rho)^2
       - \frac{1}{16} \left[ 3 t_1 \left(\frac{1}{2}+x_1\right)
                   + t_2 \left(\frac{1}{2}+x_2\right) \right]
              [(\vec\nabla\rho_p)^2 + (\vec\nabla\rho_n)^2]
              \nonumber  \\
   &=& - \frac{1}{16} \left[ 3 t_1 \left(1 + \frac{x_1}{2}\right)
                   - t_2 \left(1 + \frac{x_2}{2}\right) \right]
                \rho\nabla^2\rho
       + \frac{1}{16} \left[ 3 t_1 \left(\frac{1}{2}+x_1\right)
                   + t_2 \left(\frac{1}{2}+x_2\right) \right]
              (\rho_p\nabla^2\rho_p + \rho_n\nabla^2\rho_n)
               \nonumber  \\
 H_C &=& \frac{e^2}{2} \rho_p(\vec r) 
              \int d^3 r' \frac{\rho_p(\vec r')}{|\vec r - \vec r'|}
         - \frac{3e^2}{4} \left(\frac{3}{\pi}\right)^{1/3} \rho_p^{4/3}(\vec r)
       \label{hamilt}
\end{eqnarray}
The $H(\vec r)$ has a bulk part $H_B(\vec r)$, a surface part $H_S(\vec r)$
with gradient terms and a Coulomb term $H_C(\vec r)$. 
Here $\tau_q = \sum_{j \epsilon q} |i \vec\nabla \psi_j|^2 
   = \int \frac{p^2}{\hbar^2} f_q(\vec r, \vec p) d^3p$
and $\rho_q = \sum_{j\epsilon q} |\psi_j|^2 = \int f_q(\vec r, \vec p) d^3 p$. 
The gradient terms in Eq.(\ref{hamilt}) are important in finite nuclei 
and the Coulomb term is important for the charged proton component.  
The $t_0$, $t_1$, $t_2$, $t_3$ and $x_0$, $x_1$, $x_2$, $x_3$ are 
parameters. Different choices of these parameters give rise to 
different Skyrme interactions. 

The effective mass $m_q^*$ is
\begin{eqnarray}
 \frac{m}{m_q^*} &=& 1 + \frac{2 m}{\hbar^2} \left\{
        \frac{1}{4} \left[ t_1 \left(1 + \frac{x_1}{2}\right) 
           + t_2 \left(1 + \frac{x_2}{2}\right) \right] \rho  
      - \frac{1}{4}\left[ t_1 \left(\frac{1}{2} + x_1\right)
           - t_2 \left(\frac{1}{2} + x_2\right) \right] \rho_q \right\} 
                 \nonumber  \\
  &=& 1 + \frac{2 m}{\hbar^2} \left\{
       \frac{1}{16} \left[3 t_1 + (5 + 4 x_2) t_2\right] \rho
      \mp \frac{1}{8} \left[ t_1 \left(\frac{1}{2} + x_1\right)
           - t_2 \left(\frac{1}{2} + x_2\right) \right]
         \rho (2y-1) \right\}
\end{eqnarray}
where $q = n, p$ for neutron or proton.
At low $T$, degenerated Fermi gas model gives 
\begin{eqnarray}
 \tau_q(\vec r) &=& \frac{3}{5} \left(\frac{6\pi^2}{\gamma}\right)^{2/3}
   \left[\rho_q^{5/3}
   + \frac{5\pi^2 m_q^{*2}}{3\hbar^4} \left(\frac{\gamma}{6\pi^2}\right)^{4/3} 
      \rho_q^{1/3} T^2 + \cdots \right]    \label{tauq}
\end{eqnarray}
At zero $T$, extended Thomas Fermi approximation gives
\begin{eqnarray}
 \tau_q(\vec r) 
  &=& \frac{3}{5} \left(\frac{6\pi^2}{\gamma}\right)^{2/3} \rho_q^{5/3}
      + \frac{1}{36} \frac{(\nabla\rho_q)^2}{\rho_q}
      + \frac{1}{3} \nabla^2\rho_q       \label{tauetf}
\end{eqnarray}
Since the density gradient term depend on the slope at surface
region the ratio of surface to volume kinetic energy might be
sensitive to this term.
For a Fermi density, the integral of the last term is zero 
while the ratio of the integral of the second term without the numerical
factor (1/36) to the integral of the first term including all factors is
\begin{eqnarray}
 \frac{\left(\frac{\rho_{qc}}{2Ra}\right) \left[3 + 6 \left(\frac{a}{R_q}\right)
           + \pi^2 \left(\frac{a}{R_q}\right)^2\right] }
      {\frac{3}{5} \left(\frac{6\pi^2}{\gamma}\right)^{2/3} \rho_{qc}^{5/3}  
        \left[1 - 2.28 \left(\frac{a}{R_q}\right)
           + 9.10 \left(\frac{a}{R_q}\right)^2
           - 7.81 \left(\frac{a}{R_q}\right)^3\right]}    \nonumber
\end{eqnarray}
which is the order of one. 
Thus the second term of Eq.(\ref{tauetf}) is only a few percent of 
the first term because of the 1/36 factor.
Furthermore, this term is $T$ independent and independent of
neutron skin size since it depends only on the slope. 
Thus the density gradient term would not affect
much the relative effect of the $T$ dependent to $T$ independent part
and also the relative effect of neutron skin size $t$ to $t$ independent 
part. Since we are more interested in the temperature dependence
and the neutron skin size dependence
of energy expansion coefficients we neglect the gradient dependent terms
of Eq.(\ref{tauetf}) here in evaluation of total nuclear energy.
It is  shown that the gradient dependent correction to the coefficient 
of $T^2$ term modifies numerical results only little \cite{pr123,npa462}.

\section{Expansion of density}  \label{appenrho}

The central density $\rho_{qc}$ of the density distribution for a given 
value of $R_q$ should be determined to give a fixed number of nucleons $N_q$;
\begin{eqnarray}
 N_q(R_q) &=& \int d^3 r \rho_q(\vec r)
    = 4\pi \int_0^\infty r^2 dr 
         \frac{\rho_{qc}(R_q)}{1 + e^{(r - R_q)/a}}
    = \rho_{qc}(R_q) \frac{4\pi}{3} R_q^3 \left[1
           + \pi^2 \left(\frac{a}{R_q}\right)^2\right]    \label{nqnorm}
\end{eqnarray}
This normaization condition gives up to first order in $dR_q = (R_q - R)$,
\begin{eqnarray}
 dN_q &=& \frac{d\rho_{qc}}{dR} \frac{4\pi}{3} R^3 \left[1
           + \pi^2 \left(\frac{a}{R}\right)^2\right] d R_q 
        + \rho_{qc} \frac{4\pi}{3} R^2 \left[3
           + \pi^2 \left(\frac{a}{R}\right)^2\right] d R_q
     = 0     \\
 \frac{d\rho_{qc}}{dR} &=& - \frac{\rho_{qc}(R)}{R} 
        \frac{\left[3 + \pi^2 \left(\frac{a}{R}\right)^2\right]}
             {\left[1 + \pi^2 \left(\frac{a}{R}\right)^2\right]}  \\
 \rho_{qc}(R_q) &=& \rho_{qc}(R) \left[1
      - \left(\frac{3 + \pi^2 \left(\frac{a}{R}\right)^2}
                   {1 + \pi^2 \left(\frac{a}{R}\right)^2}\right)
        \left(\frac{R_q - R}{R}\right) + \cdots \right]    \label{rhoqc}
\end{eqnarray}
Thus the first order correction coefficient to $\rho_{qc}^m$ due to 
fixed $N_q$ is the zeroth order term times
 $-\frac{m}{R} \left[\frac{3 + \pi^2 \left(\frac{a}{R}\right)^2}
                   {1 + \pi^2 \left(\frac{a}{R}\right)^2}\right]$.
This result is independent of which type of particle, neutron or proton.

If we keep the particle number $N_q$ and the total central density
$\rho_c = \rho_{nc} + \rho_{pc}$ to be a constant while varying $R_q$, then
\begin{eqnarray}
 \rho_c(R) &=& \rho_{nc}(R_n) + \rho_{pc}(R_p)     \nonumber  \\
   &=& \rho_{nc}(R) + \rho_{pc}(R)
     - \left[\frac{3 + \pi^2 \left(\frac{a}{R}\right)^2}
                     {1 + \pi^2 \left(\frac{a}{R}\right)^2}\right]
          \left[\rho_{nc}(R) \left(\frac{R_n - R}{R}\right)
            + \rho_{pc}(R) \left(\frac{R_p - R}{R}\right)\right]
              \nonumber  \\
   &=& \rho_{nc}(R) + \rho_{pc}(R)
\end{eqnarray}
Thus we get
\begin{eqnarray}
 \rho_{nc} \left(\frac{R_n - R}{R}\right)
    + \rho_{pc} \left(\frac{R_p - R}{R}\right) = 0
\end{eqnarray}
and
\begin{eqnarray}
 R_p - R &=& - \frac{\rho_{nc}}{\rho_{pc}} (R_n - R)     \nonumber  \\
 t = R_n - R_p &=& (R_n - R) - (R_p - R)
   = (R_n - R) \left(1 + \frac{\rho_{nc}}{\rho_{pc}}\right)
   = \frac{\rho_c}{\rho_{pc}} (R_n - R)
\end{eqnarray}
Finally we have 
\begin{eqnarray}
 t_n &=& R_n - R = \frac{\rho_{pc}}{\rho_c} t ,      \nonumber  \\
 t_p &=& R_p - R = - \frac{\rho_{nc}}{\rho_c} t      \label{tqtrho}
\end{eqnarray}
to lowest order in $t$.
The same result can be obtained by requiring $A = N + Z$ constant
with keeping the central densities $\rho_{qc}$ unchanged 
in the $dR_q$ expansion.
When $R_q = R$, $\rho_{qc}(R)/\rho_c(R) = N_q/A$ and thus we get
\begin{eqnarray}
 t_n &=& R_n - R = \frac{Z}{A} t ,      \nonumber  \\
 t_p &=& R_p - R = - \frac{N}{A} t      \label{tqtap}
\end{eqnarray}
This is the same result given in Ref.\cite{dan16}.
Due to Eq.(\ref{tqtrho}) or (\ref{tqtap}), $t_n > 0$ and $t_p < 0$ 
with $t > 0$ for $N > Z$
while $t_n < 0$ and $t_p > 0$ with $t < 0$ for $Z > N$.
Thus the role of $t_n$ and $t_p$ is exchanged as the sign of $t$ changes.

Since the size $R_q$ depends on the central density $\rho_{qc}$ for a 
given value of particle number $N_q$ as in Eq.(\ref{nqnorm}),
the neutron skin size $t$ is related to the particle number $N_q$ 
and the central density $\rho_{qc}$.
Using Eqs.(\ref{nqnorm}) and (\ref{tqtrho}) we can obtain following
conditions to lowest order in $t$.
\begin{eqnarray}
 N - Z &=& \frac{4\pi}{3} \left[\left(R_n^3 + \pi^2 a^2 R_n\right) \rho_{nc}
                         - \left(R_p^3 + \pi^2 a^2 R_p\right) \rho_{pc}\right]
            \nonumber  \\
  &=& \frac{4\pi}{3} \left[
      R^3 \left(1 + \frac{\rho_{pc}}{\rho_c} \frac{t}{R}\right)^3 \rho_{nc}
    + \pi^2 a^2 R \left(1 + \frac{\rho_{pc}}{\rho_c} \frac{t}{R}\right)\rho_{nc}
               \right.   \nonumber  \\  & &  \left.
    - R^3 \left(1 - \frac{\rho_{nc}}{\rho_c} \frac{t}{R}\right)^3 \rho_{pc}
    - \pi^2 a^2 R \left(1 - \frac{\rho_{nc}}{\rho_c} \frac{t}{R}\right)
                \rho_{pc} \right]
            \nonumber  \\
  &\approx& \frac{4\pi}{3} \left[
      R^3 \left(1 + 3 \frac{\rho_{pc}}{\rho_c} \frac{t}{R}\right) \rho_{nc}
    - R^3 \left(1 - 3 \frac{\rho_{nc}}{\rho_c} \frac{t}{R}\right) \rho_{pc}
               \right.   \nonumber  \\  & &  \left.
    + \pi^2 a^2 R \left(1 + \frac{\rho_{pc}}{\rho_c} \frac{t}{R}\right)\rho_{nc}
    - \pi^2 a^2 R \left(1 - \frac{\rho_{nc}}{\rho_c} \frac{t}{R}\right) 
                \rho_{pc} \right]
            \nonumber  \\
  &=& \frac{4\pi}{3} \left(R^3 + \pi^2 a^2 R\right) 
                     \left(\rho_{nc} - \rho_{pc}\right)
     + \frac{4\pi}{3} \left(3 R^3 + \pi^2 a^2 R\right)
            \left(\frac{\rho_{nc}}{\rho_c} \rho_{pc}
                + \frac{\rho_{pc}}{\rho_c} \rho_{nc}\right) \frac{t}{R}  
            \nonumber  \\
  &=& \frac{4\pi}{3} R^3 \left[1 + \pi^2 \left(\frac{a}{R}\right)^2\right]
           \rho_c \left(\frac{\rho_{nc} - \rho_{pc}}{\rho_c}\right)
    + \frac{4\pi}{3} \left[R^3 + \frac{\pi^2 a^2}{3} R\right] 
           3 \left(\frac{\rho_{nc}}{\rho_c} \rho_{pc}
                + \frac{\rho_{pc}}{\rho_c} \rho_{nc}\right) \frac{t}{R}
            \nonumber  \\
  &=& A \left(\frac{\rho_{nc} - \rho_{pc}}{\rho_c}\right)
    + 3 \frac{A}{\rho_c} \left(\frac{\rho_{nc}}{\rho_c} \rho_{pc}
                + \frac{\rho_{pc}}{\rho_c} \rho_{nc}\right)  
       \left[\frac{1 + \frac{1}{3} \left(\frac{\pi a}{R}\right)^2}
                  {1 + \left(\frac{\pi a}{R}\right)^2}\right]
         \frac{t}{R}      \\
 \frac{t}{R} &=& \frac{1}{3} \frac{1}{A} 
        \frac{\rho_c}{\left(\frac{\rho_{nc}}{\rho_c} \rho_{pc}
                + \frac{\rho_{pc}}{\rho_c} \rho_{nc}\right)}
        \left[\left(N - Z\right) 
          - A \left(\frac{\rho_{nc} - \rho_{pc}}{\rho_c}\right)\right]
       \left[\frac{1 + \left(\frac{\pi a}{R}\right)^2}
                  {1 + \frac{1}{3} \left(\frac{\pi a}{R}\right)^2}\right]
               \nonumber  \\
  &=& \frac{1}{3} \left(\frac{1}{2 (1-y_c) y_c}\right)
      \left[\left(\frac{N - Z}{A}\right) - (1 - 2y_c)\right]
       \left[\frac{1 + \left(\frac{\pi a}{R}\right)^2}
                  {1 + \frac{1}{3} \left(\frac{\pi a}{R}\right)^2}\right]
           \label{trrho}
\end{eqnarray}
where $y_c = \rho_{pc}/\rho_c$ is the proton fraction of the central density.
For the case of $\rho_{nc} \approx \rho_{pc}$, $y_c = 1/2 + \epsilon$ 
and $(1 - y_c) = 1/2 - \epsilon$.
Then the factor $[2(1-y_c)y_c]^{-1}$ becomes
\begin{eqnarray}
 \frac{1}{2(1-y_c)y_c} &=& \frac{2}{(1 - 2 \epsilon)(1 + 2 \epsilon)} 
    \approx 2 (1 - 4 \epsilon^2)    
\end{eqnarray}
Thus for an uniform distribution (diffuseness parameter $a = 0$)
with $\rho_{nc} \approx \rho_{pc}$, 
the neutron skin size $t/R$ of Eq.(\ref{trrho}) becomes, 
up to first order in $\epsilon$, 
\begin{eqnarray}
 \frac{t}{R} &=& \frac{2}{3} \frac{N-Z}{A} - \frac{2}{3} (1 - 2 y_c)
                        \label{trmeyer}
\end{eqnarray}
which is the result given in Ref.\cite{dan16}.
It is shown that the empirical neutron skin size $t$ is approximately 
proportional to $I = (N-Z)/A$ \cite{skinprl87}.  
On the other hand, as another form,
\begin{eqnarray}
 \frac{N}{\rho_{nc}} - \frac{Z}{\rho_{pc}}
  &=& \frac{4\pi}{3} \left[\left(R_n^3 + \pi^2 a^2 R_n\right)
                         - \left(R_p^3 + \pi^2 a^2 R_p\right)\right]
            \nonumber  \\
  &=& \frac{4\pi}{3} \left[
      R^3 \left(1 + \frac{\rho_{pc}}{\rho_c} \frac{t}{R}\right)^3
    - R^3 \left(1 - \frac{\rho_{nc}}{\rho_c} \frac{t}{R}\right)^3
    + \pi^2 a^2 R \left(1 + \frac{\rho_{pc}}{\rho_c} \frac{t}{R}\right)
    - \pi^2 a^2 R \left(1 - \frac{\rho_{nc}}{\rho_c} \frac{t}{R}\right)\right]
            \nonumber  \\
  &\approx& \frac{4\pi}{3} \left[
      R^3 \left(1 + 3 \frac{\rho_{pc}}{\rho_c} \frac{t}{R}\right)
    - R^3 \left(1 - 3 \frac{\rho_{nc}}{\rho_c} \frac{t}{R}\right)
    + \pi^2 a^2 R \left(1 + \frac{\rho_{pc}}{\rho_c} \frac{t}{R}\right)
    - \pi^2 a^2 R \left(1 - \frac{\rho_{nc}}{\rho_c} \frac{t}{R}\right)\right]
            \nonumber  \\
  &=& \frac{4\pi}{3} \left[R^3 3 \left(\frac{\rho_{pc}}{\rho_c} 
                          + \frac{\rho_{nc}}{\rho_c}\right)
              + \pi^2 a^2 R \left(\frac{\rho_{pc}}{\rho_c} 
                          + \frac{\rho_{nc}}{\rho_c}\right)\right]
            \frac{t}{R}  
   = \frac{4\pi}{3} 3 \left[R^3 + \frac{\pi^2 a^2}{3} R\right] \frac{t}{R}
           \nonumber  \\
  &=& 3 \frac{A}{\rho_c} 
       \left[\frac{1 + \frac{1}{3} \left(\frac{\pi a}{R}\right)^2}
                  {1 + \left(\frac{\pi a}{R}\right)^2}\right]
         \frac{t}{R}      \\
 \frac{t}{R} &=& \frac{1}{3} \frac{\rho_c}{A} 
        \left(\frac{N}{\rho_{nc}} - \frac{Z}{\rho_{pc}}\right)
       \left[\frac{1 + \left(\frac{\pi a}{R}\right)^2}
                  {1 + \frac{1}{3} \left(\frac{\pi a}{R}\right)^2}\right]
   = \frac{1}{3} \frac{1}{A} \left(\frac{N}{1-y_c} - \frac{Z}{y_c}\right)
       \left[\frac{1 + \left(\frac{\pi a}{R}\right)^2}
                  {1 + \frac{1}{3} \left(\frac{\pi a}{R}\right)^2}\right]
                    \label{trrhob}
\end{eqnarray}
Even if Eqs.(\ref{trrho}) and (\ref{trrhob}) look different 
they are the same equation.
The proton ratio $y_c$ has the range of $Z/A \le y_c \le 1/2$ for finite 
nuclei where $y_c = Z/A$ when $R_n = R_p$ and $y_c = 1/2$ 
for $\rho_{nc} = \rho_{pc}$.

For one extreme case of the same central density $\rho_{nc} = \rho_{pc}$,
the proton ratio $y = 1/2$ and 
\begin{eqnarray}
 \frac{t}{R} &=& \frac{2}{3} \left(\frac{N - Z}{A}\right)
       \left[\frac{1 + \left(\frac{\pi a}{R}\right)^2}
                  {1 + \frac{1}{3} \left(\frac{\pi a}{R}\right)^2}\right]
   = \frac{2}{3} \left[\frac{1 + \left(\frac{\pi a}{R}\right)^2}
           {1 + \frac{1}{3} \left(\frac{\pi a}{R}\right)^2}\right] I
\end{eqnarray}
Thus the neutron skin size is linearly proportional to the isospin
factor $I$ when the proton and neutron central densities are same.
For the other extreme case of the same size $R_n = R_p = R$, 
the proton ratio is $y_c = Z/A$ with $1-y_c = N/A$, and thus 
the neutron skin size becomes $t/R = 0$.
However when $R_n \approx R_p$ with $y_c = Z/A + \epsilon$, 
we have $1 - y_c = N/A - \epsilon$ and, from Eq.(\ref{trrho}),
\begin{eqnarray}
 \frac{t}{R} &=& \frac{1}{3} 
    \left[\frac{1}{2 (N/A - \epsilon) (Z/A + \epsilon)}\right]
    \left[\left(\frac{N - Z}{A}\right) - \left(\frac{N}{A} - \epsilon
           - \frac{Z}{A} - \epsilon\right)\right]
       \left[\frac{1 + \left(\frac{\pi a}{R}\right)^2}
                  {1 + \frac{1}{3} \left(\frac{\pi a}{R}\right)^2}\right]
           \nonumber  \\
  &\approx&  \frac{1}{3} \left[\frac{1}{2 (N/A) (Z/A)}\right]
     \left(1 + \frac{A}{N} \epsilon\right) \left(1 - \frac{A}{Z} \epsilon\right)
    \left[\left(\frac{N - Z}{A}\right) - \left(\frac{N - Z}{A}\right) 
          + 2 \epsilon\right]
       \left[\frac{1 + \left(\frac{\pi a}{R}\right)^2}
                  {1 + \frac{1}{3} \left(\frac{\pi a}{R}\right)^2}\right]
           \nonumber  \\
  &\approx& \frac{1}{3} \left(\frac{A^2}{N Z}\right) 
    \left[\frac{1 + \left(\frac{\pi a}{R}\right)^2}
         {1 + \frac{1}{3} \left(\frac{\pi a}{R}\right)^2}\right] \epsilon
   =  \frac{1}{3} \left(\frac{4}{1 - I^2}\right)
    \left[\frac{1 + \left(\frac{\pi a}{R}\right)^2}
         {1 + \frac{1}{3} \left(\frac{\pi a}{R}\right)^2}\right] \epsilon
\end{eqnarray}
up to the first order in $\epsilon$.

Using Eq.(\ref{rhoqc}), the Fermi density Eq.(\ref{fermden}) is expanded 
up to first order in $(R_q - R)$ as
\begin{eqnarray}
 \rho_q(r) &=& \frac{\rho_{qc}(R_q)}{1 + e^{(r-R_q)/a}}
     = \frac{\rho_{qc}(R_q)}{1 + e^{(y + (R-R_q)/a)}}   \nonumber  \\
   &=& \frac{\rho_{qc}(R)}{1 + e^y}
     \left[1 - \left(\frac{e^y}{1 + e^y}\right) \left(\frac{R - R_q}{a}
            \right) + \cdots\right]
     \left[1 - \left(\frac{3 + \pi^2 \left(\frac{a}{R}\right)^2}
                   {1 + \pi^2 \left(\frac{a}{R}\right)^2}\right)
         \left(\frac{R_q - R}{R}\right) + \cdots \right]   \nonumber  \\
   &=& \left(\frac{\rho_{qc}(R)}{1 + e^y}\right) \left[1
          - \left(\frac{e^y}{1 + e^y}\right) \left(\frac{R - R_q}{a}\right)
          - \left(\frac{3 + \pi^2 \left(\frac{a}{R}\right)^2}
                   {1 + \pi^2 \left(\frac{a}{R}\right)^2}\right)
              \left(\frac{R_q - R}{R}\right) + \cdots \right]  \nonumber  \\
   &=& \left(\frac{\rho_{qc}(R)}{1 + e^y}\right) \left[1
          + \left(\frac{e^y}{1 + e^y} \frac{R}{a}
                - \frac{3 + \pi^2 \left(\frac{a}{R}\right)^2}
                   {1 + \pi^2 \left(\frac{a}{R}\right)^2}\right)
              \left(\frac{R_q - R}{R}\right) + \cdots \right] \label{rhoqr} \\
 \rho_q^m(r) &=& \left(\frac{\rho_{qc}(R)}{1 + e^y}\right)^m \left[1
          + m \left(\frac{e^y}{1 + e^y} \frac{R}{a}
               - \frac{3 + \pi^2 \left(\frac{a}{R}\right)^2}
                   {1 + \pi^2 \left(\frac{a}{R}\right)^2}\right)
              \left(\frac{R_q - R}{R}\right) + \cdots \right]   
\end{eqnarray}
where $y = (r - R)/a$.
Since $t_n = \frac{Z}{A} t$ and $t_p = - \frac{N}{Z} t$ (Eq.(\ref{tqtap})) 
with $\rho_{qc}(R)/\rho_c(R) = N_q/A$, 
\begin{eqnarray}
 \rho_n^m(r) + \rho_p^m(R)
   &=& \left(\frac{\rho_c(R)}{1 + e^{(r-R)/a}}\right)^m
       \left[\left(\frac{\rho_{nc}^m(R) + \rho_{pc}^m(R)}{\rho_c^m(R)}\right)
            \right.   \nonumber  \\  & & \hspace{0.5in}  \left.
        + m \left(\frac{e^y}{1+e^y} \frac{R}{a}
             - \frac{3 + \pi^2 \left(\frac{a}{R}\right)^2}
                   {1 + \pi^2 \left(\frac{a}{R}\right)^2}\right)
             \left(\frac{\rho_{nc}^m(R) Z - \rho_{pc}^m(R) N}{\rho_c^m(R) A}
             \right) \frac{(R_n - R_p)}{R}
        + \cdots \right]     \nonumber  \\
   &\approx& \left(\frac{\rho_c(R)}{1 + e^y}\right)^m
     \left[\left(\frac{N^m + Z^m}{A^m}\right)
        + m \left(\frac{e^y}{1+e^y} \frac{R}{a}
             - \frac{3 + \pi^2 \left(\frac{a}{R}\right)^2}
                   {1 + \pi^2 \left(\frac{a}{R}\right)^2}\right)
             \left(\frac{N^m Z - Z^m N}{A^{m+1}}\right) \frac{t}{R}
        + \cdots \right]       \label{rhomnp}
\end{eqnarray}
From this result it is easy to show that the
quantity $\rho_n^m(r) + \rho_p^m(r)$ has a first order correction 
from skin size $t$ and the first order correction vanishes for $m = 1$.
That is the total density $\rho(r) = \rho_n(r) + \rho_p(r)$ is independent
of the neutron skin size $t = R_n - R_p$ up to first order.
Thus only the explicit $\rho_q$ dependent terms in Skyrme interaction,
not the total density $\rho$ dependent terms, depend on the skin size $t$
up to the first order.

\section{Integral of density functional} \label{appendc}

Since $F[\rho_q] = \int d^3r f(\rho_q(\vec r))$ for a Feremi density
 $\rho_q(r) = \frac{\rho_{qc}}{1 + e^{(r-R_q)/a}}$,
where $f(\rho_q)$ is a function of a single density $\rho_q$ only,
can be integrated exactly as a function $F(R_q)$ of $R_q$ \cite{prc82}, 
we can expand $F(R_q)$ easily
in terms of $t_q = R_q - R$ or $x_q = - t_q/a$ around $F(R)$.
That is
\begin{eqnarray}
 F(R_q) &=& F(R) + \frac{d F(R)}{d R} t_q
           + \frac{d^2 F(R)}{d R^2} \frac{t_q^2}{2} + \cdots
\end{eqnarray}
This is much simpler than using preveous method of expanding
in $t$ (Eq.(\ref{rhoqr})) first then integrate the results.
For fixed $N_q$, the central density $\rho_{qc}^m(R_q)$ is expanded as
\begin{eqnarray}
 \rho_{qc}^m(R_q) &=& \rho_{qc}^m(R) \left[1
    - m \left(\frac{3 + \pi^2 \left(\frac{a}{R}\right)^2}
                   {1 + \pi^2 \left(\frac{a}{R}\right)^2}\right)
         \left(\frac{R_q - R}{R}\right) + \cdots \right]
\end{eqnarray}
Thus
\begin{eqnarray}
 \int d^3 r \rho_q^2 (\vec r)
  &=& \frac{4\pi}{3} \rho_{qc}^2(R_q)
        \left(R_q^3 - 3 a R_q^2 + \pi^2 a^2 R_q - \pi^2 a^3\right)
                \nonumber  \\
  &=& \frac{4\pi}{3} R^3 \rho_{qc}^2(R) \left[1 - 3 \left(\frac{a}{R}\right)
           + \pi^2 \left(\frac{a}{R}\right)^2
           - \pi^2 \left(\frac{a}{R}\right)^3\right] \left[1
    - 2 \left(\frac{3 + \pi^2 \left(\frac{a}{R}\right)^2}
                   {1 + \pi^2 \left(\frac{a}{R}\right)^2}\right)
         \left(\frac{R_q - R}{R}\right)\right]           \nonumber  \\
  & & + \frac{4\pi}{3} R^3 \rho_{qc}^2(R) \frac{a}{R}
        \left[3 - 6 \left(\frac{a}{R}\right)
            + \pi^2 \left(\frac{a}{R}\right)^2\right]
        \left(\frac{R_q - R}{a}\right) + \cdots           \\
 \int d^3 r \rho_q(\vec r) \nabla^2 \rho_q(\vec r)
  &=& - \frac{4\pi}{3} R_q^3 \frac{\rho_{qc}^2(R_q)}{2R_q a} \left[1
     + \left(\frac{\pi^2}{3} - 2\right) \left(\frac{a}{R_q}\right)^2\right]
              \nonumber  \\
  &=& - \frac{4\pi}{3} R^3 \frac{\rho_{qc}^2(R)}{2R a} \left[1
         + \left(\frac{\pi^2}{3} - 2\right) \left(\frac{a}{R}\right)^2\right]
       \left[1 - 2 \left(\frac{3 + \pi^2 \left(\frac{a}{R}\right)^2}
                   {1 + \pi^2 \left(\frac{a}{R}\right)^2}\right)
         \left(\frac{R_q - R}{R}\right)\right]           \nonumber  \\
  & & - \frac{4\pi}{3} R^3 \frac{\rho_{qc}^2(R)}{R^2}
        \left(\frac{R_q - R}{a}\right) + \cdots           \\
 \int d^3 r \rho_q^{4/3}(\vec r)
  &=& \frac{4\pi}{3} R_q^3 \rho_{qc}^{4/3}(R_q) \left[1
      - 1.335546875 \left(\frac{a}{R_q}\right)
      + 8.81615625 \left(\frac{a}{R_q}\right)^2
      - 5.0303125 \left(\frac{a}{R_q}\right)^3\right]     \nonumber  \\
  &=& \frac{4\pi}{3} R^3 \rho_{qc}^{4/3}(R) \left[1
      - 1.335546875 \left(\frac{a}{R}\right)
      + 8.81615625 \left(\frac{a}{R}\right)^2
      - 5.0303125 \left(\frac{a}{R}\right)^3\right]
                \nonumber  \\   & &    \hspace{0.5in} \times
       \left[1 - \frac{4}{3} \left(\frac{3 + \pi^2 \left(\frac{a}{R}\right)^2}
                   {1 + \pi^2 \left(\frac{a}{R}\right)^2}\right)
         \left(\frac{R_q - R}{R}\right)\right]
                \nonumber  \\   & &
    + \frac{4\pi}{3} R^3 \rho_{qc}^{4/3}(R) \frac{a}{R} \left[3
      - 2.67109375 \left(\frac{a}{R}\right)
      + 8.81615625 \left(\frac{a}{R}\right)^2\right]
        \left(\frac{R_q - R}{a}\right) + \cdots      \label{ecex}  \\
 \int d^3 r \rho_p^{5/3}(\vec r)
  &=& \frac{4\pi}{3} R_q^3 \rho_{qc}^{5/3}(R_q) \left[1
      - 2.276943 \left(\frac{a}{R_q}\right)
      + 9.10458 \left(\frac{a}{R_q}\right)^2
      - 7.80506 \left(\frac{a}{R_q}\right)^3\right]     \nonumber  \\
  &=& \frac{4\pi}{3} R^3 \rho_{qc}^{5/3}(R) \left[1
      - 2.276943 \left(\frac{a}{R}\right)
      + 9.10458 \left(\frac{a}{R}\right)^2
      - 7.80506 \left(\frac{a}{R}\right)^3\right]
                \nonumber  \\   & &   \hspace{0.5in} \times
       \left[1 - \frac{5}{3} \left(\frac{3 + \pi^2 \left(\frac{a}{R}\right)^2}
                   {1 + \pi^2 \left(\frac{a}{R}\right)^2}\right)
         \left(\frac{R_q - R}{R}\right)\right]
                \nonumber  \\   & &
    + \frac{4\pi}{3} R^3 \rho_{qc}^{5/3}(R) \frac{a}{R} \left[3
      - 4.553886 \left(\frac{a}{R}\right)
      + 9.10458 \left(\frac{a}{R}\right)^2\right]
        \left(\frac{R_q - R}{a}\right) + \cdots           \\
 \int d^3 r \rho_p^{8/3}(\vec r)
  &=& \frac{4\pi}{3} R_q^3 \rho_{qc}^{8/3}(R_q) \left[1
      - 4.07693333 \left(\frac{a}{R_q}\right)
      + 11.836907 \left(\frac{a}{R_q}\right)^2
      - 13.26781 \left(\frac{a}{R_q}\right)^3\right]     \nonumber  \\
  &=& \frac{4\pi}{3} R^3 \rho_{qc}^{8/3}(R) \left[1
      - 4.07693333 \left(\frac{a}{R}\right)
      + 11.836907 \left(\frac{a}{R}\right)^2
      - 13.26781 \left(\frac{a}{R}\right)^3\right]
                \nonumber  \\   & &    \hspace{0.5in} \times
       \left[1 - \frac{8}{3} \left(\frac{3 + \pi^2 \left(\frac{a}{R}\right)^2}
                   {1 + \pi^2 \left(\frac{a}{R}\right)^2}\right)
         \left(\frac{R_q - R}{R}\right)\right]
                \nonumber  \\   & &
    + \frac{4\pi}{3} R^3 \rho_{qc}^{8/3}(R) \frac{a}{R} \left[3
      - 8.15386666 \left(\frac{a}{R}\right)
      + 11.836907 \left(\frac{a}{R}\right)^2\right]
        \left(\frac{R_q - R}{a}\right) + \cdots           \\
 E_C &=& \frac{3}{5} \frac{Z^2 e^2}{R_p} \left[1
      - \left(\frac{7 \pi^2}{6}\right) \left(\frac{a}{R_p}\right)^2\right]
              \nonumber  \\
  &=& \frac{3}{5} \frac{Z^2 e^2}{R} \left[1
      - \left(\frac{7 \pi^2}{6}\right) \left(\frac{a}{R}\right)^2\right]
    - \frac{3}{5} \frac{Z^2 e^2}{R} \frac{a}{R} \left[1
      - \left(\frac{7 \pi^2}{2}\right) \left(\frac{a}{R}\right)^2\right]
       \left(\frac{R_p - R}{a}\right) + \cdots      \label{ecoul}
\end{eqnarray}
In Ref.\cite{prc82}, the Coulomb exchange term is shown only upto 0th order 
in $a/R$ but the actual calculation included all terms of $a/R$ up to 3.
For the term with mixed densities, we need to integrate after expansion.
Up to 1st order in $x = -t/a$,
\begin{eqnarray}
 \int d^3 r \rho_q^m (\vec r) \rho^\alpha(\vec r)
  &=& 4\pi \rho_{qc}^m(R_q) \rho_c^\alpha(R) \int_0^\infty r^2 dr
      \left(\frac{1}{1 + e^{(r-R_q)/a}}\right)^m
      \left(\frac{\rho_{nc}(R_n)/\rho_c(R)}{1 + e^{(r-R_n)/a}}
          + \frac{\rho_{pc}(R_p)/\rho_c(R)}{1 + e^{(r-R_p)/a}}\right)^\alpha
             \nonumber \\
  &\approx& 4\pi \rho_{qc}^m(R) \rho_c^\alpha(R) \int_0^\infty r^2 dr
      \left(\frac{1}{1 + e^{(r-R)/a}}\right)^{\alpha+m}
       \left[1 - m \left(\frac{3 + \pi^2 \left(\frac{a}{R}\right)^2}
                   {1 + \pi^2 \left(\frac{a}{R}\right)^2}\right)
         \left(\frac{R_q - R}{R}\right)\right]
                \nonumber  \\   & &
    - 4\pi \rho_{qc}^m(R) \rho_c^\alpha(R) \int_0^\infty r^2 dr
      \left(\frac{1}{1 + e^{(r-R)/a}}\right)^{\alpha+m}
      \left(\frac{e^{(r-R)/a}}{1 + e^{(e-R)/a}}\right)
      m \left(\frac{R-R_q}{a}\right)    \nonumber \\
  &=& \frac{4\pi}{3} \rho_{qc}^m(R) \rho_c^\alpha(R)  \int_{-\infty}^\infty dy
      (ay + R)^3 \frac{(\alpha+m) e^y}{(1 + e^y)^{\alpha+m+1}}
       \left[1 - m \left(\frac{3 + \pi^2 \left(\frac{a}{R}\right)^2}
                   {1 + \pi^2 \left(\frac{a}{R}\right)^2}\right)
         \left(\frac{R_q - R}{R}\right)\right]
                \nonumber  \\   & &
    + 4 m \pi \rho_{qc}^m(R) \rho_c^\alpha(R) a \int_{-\infty}^\infty dy
      (ay + R)^2 \frac{e^y}{(1 + e^y)^{\alpha+m+1}}
      \left(\frac{R_q-R}{a}\right)
\end{eqnarray}
For SLy4 parameter with $\alpha = 1/6$ and $m = 2$,
\begin{eqnarray}
 \int d^3 r \rho_q^2 (\vec r) \rho^{1/6}(\vec r)
  &=& \frac{4\pi}{3} R^3 \rho_{qc}^2 \rho_c^{1/6} \left[1
       - 3.30669 \left(\frac{a}{R}\right) + 10.331 \left(\frac{a}{R}\right)^2
       - 10.7804 \left(\frac{a}{R}\right)^3\right]
             \nonumber  \\  & &    \hspace{0.5in} \times
       \left[1 - 2 \left(\frac{3 + \pi^2 \left(\frac{a}{R}\right)^2}
                   {1 + \pi^2 \left(\frac{a}{R}\right)^2}\right)
         \left(\frac{R_q - R}{R}\right)\right]
                \nonumber  \\   & &
    + \frac{4\pi}{3} R^3 \rho_{qc}^2 \rho_c^{1/6} \frac{36}{13} \frac{a}{R}
      \left[1 - 2.204466 \left(\frac{a}{R}\right)
          + 3.4436602 \left(\frac{a}{R}\right)^2\right]
      \left(\frac{R_q-R}{a}\right)
\end{eqnarray}
Here $36/13 = 3 \times 2 / (13/6)$.
For SkM($m^*=m$) parameter with $\alpha = 1$ and $m = 2$,
\begin{eqnarray}
 \int d^3 r \rho_q^2 (\vec r) \rho(\vec r)
  &=& \frac{4\pi}{3} R^3 \rho_{qc}^2 \rho_c \left[1
       - \frac{9}{2} \left(\frac{a}{R}\right)
       + (3 + \pi^2) \left(\frac{a}{R}\right)^2
       - \frac{3\pi^2}{2} \left(\frac{a}{R}\right)^3\right]
             \nonumber  \\  & &    \hspace{0.5in} \times
       \left[1 - 2 \left(\frac{3 + \pi^2 \left(\frac{a}{R}\right)^2}
                   {1 + \pi^2 \left(\frac{a}{R}\right)^2}\right)
         \left(\frac{R_q - R}{R}\right)\right]
                \nonumber  \\   & &
    + \frac{4\pi}{3} R^3 \rho_{qc}^2 \rho_c 2 \frac{a}{R}
      \left[1 - 3 \left(\frac{a}{R}\right)
       + \left(1 + \frac{\pi^2}{3}\right) \left(\frac{a}{R}\right)^2\right]
      \left(\frac{R_q-R}{a}\right)
\end{eqnarray}
Here $2 = 3 \times 2 / 3$.
For $T$-independent term in kinetic energy with $\alpha = 1$ and $m = 5/3$,
\begin{eqnarray}
 \int d^3 r \rho_q^{5/3} (\vec r) \rho(\vec r)
  &=& \frac{4\pi}{3} R^3 \rho_{qc}^{5/3} \rho_c \left[1
       - 4.07693333 \left(\frac{a}{R}\right)
       + 11.836907 \left(\frac{a}{R}\right)^2
       - 13.26781 \left(\frac{a}{R}\right)^3\right]
             \nonumber  \\  & &    \hspace{0.5in} \times
       \left[1 - \frac{5}{3} \left(\frac{3 + \pi^2 \left(\frac{a}{R}\right)^2}
                   {1 + \pi^2 \left(\frac{a}{R}\right)^2}\right)
         \left(\frac{R_q - R}{R}\right)\right]
                \nonumber  \\   & &
    + \frac{4\pi}{3} R^3 \rho_{qc}^{5/3} \rho_c \frac{15}{8} \frac{a}{R}
      \left[1 - 2.71797333 \left(\frac{a}{R}\right)
       + 3.9456266667 \left(\frac{a}{R}\right)^2\right]
      \left(\frac{R_q-R}{a}\right)
\end{eqnarray}
Here $(15/8) = 3 \times (5/3) / (8/3)$.
For kinetic energy we cannot use this method since it uses numerical
integration.

The kinetic energy has term with the form of $\rho_q^m {m_q^*}^2$ where
the effective mass has the form of $m_q/m_q^* = 1 + a \rho + b \rho_q$.
Since
\begin{eqnarray}
 \rho_q^m(r) &=& \left(\frac{\rho_{qc}(R)}{1 + e^y}\right)^m \left[1
          - m \left(\frac{e^y}{1 + e^y}\right) \left(\frac{R - R_q}{a}\right)
          - m \left(\frac{3 + \pi^2 \left(\frac{a}{R}\right)^2}
                   {1 + \pi^2 \left(\frac{a}{R}\right)^2}\right)
              \left(\frac{R_q - R}{R}\right) + \cdots \right]
\end{eqnarray}
and $\rho$ is $t$ independent up to the first order in the neutron skin $t$,
\begin{eqnarray}
 \rho_q^m {m_q^*}^n &=& \rho_q^m m_q^n (1 + a \rho + b \rho_q)^{-n}
                 \nonumber \\
  &=& \left(\frac{\rho_{qc}(R_q)}{1 + e^y}\right)^m m_q^n
        \left[1 - m \left(\frac{e^y}{1 + e^y}\right) x_q + \cdots\right]
     \left[1 + a \rho + b \left(\frac{\rho_{qc}(R_q)}{1 + e^y}\right)\left(1
        - \left(\frac{e^y}{1 + e^y}\right) x_q + \cdots\right)\right]^{-n}
                 \nonumber \\
  &=& \left(\frac{\rho_{qc}(R)}{1 + e^y}\right)^m m_q^n
      \left[1 + a \rho + b \left(\frac{\rho_{qc}(R)}{1 + e^y}\right)\right]^{-n}            \nonumber \\ & & \hspace{0.3in}\times
      \left\{1 -  \left[m   - \frac{n b \frac{\rho_{qc}(R)}{1 + e^y}}
          {\left(1 + a \rho + b \frac{\rho_{qc}(R)}{1 + e^y}\right)}\right]
           \left[\left(\frac{e^y}{1 + e^y}\right) x_q
              + \left(\frac{3 + \pi^2 \left(\frac{a}{R}\right)^2}
                   {1 + \pi^2 \left(\frac{a}{R}\right)^2}\right)
              \left(\frac{R_q - R}{R}\right)\right]
        + \cdots \right\}
                 \nonumber \\
  &=& \left.\rho_q^m {m_q^*}^n\right|_{x_q=0} \left\{1
        - \left[m - n b \rho_q \left.\frac{{m_q^*}}{m_q}\right|_{x_q=0}
         \right] \left[\left(\frac{e^y}{1 + e^y}\right) x_q
            + \left(\frac{3 + \pi^2 \left(\frac{a}{R}\right)^2}
                   {1 + \pi^2 \left(\frac{a}{R}\right)^2}\right)
              \left(\frac{R_q - R}{R}\right)\right]
        + \cdots \right\}
\end{eqnarray}
Since the $m_q^*$ factor depends on $\rho_q$, the first order term
in $x = - t/a$ of $(\rho_n^m {m_n^*}^n + \rho_p^m {m_p^*}^n)$
does not become zero for any $m$ for non-zero $n$.
Similarly total kinetic energy $\tau = \tau_n + \tau_p$ has nonzero first
order term in the neutron skin $t$ in contrast to the total density
$\rho = \rho_n + \rho_p$ which is independent to $t$ up to first order.
Total kinetic energy is
\begin{eqnarray}
 \tau &=& \frac{3}{5} \left(\frac{6\pi^2}{\gamma}\right)^{2/3}
    \left[\left(\rho_n^{5/3} + \rho_p^{5/3}\right)
      + \frac{5\pi^2}{3 \hbar^4}\left(\frac{\gamma}{6\pi^2}\right)^{4/3} T^2
         \left(\rho_n^{1/3} {m_n^*}^2 + \rho_p^{1/3} {m_p^*}^2\right)
      + \cdots \right]          \nonumber  \\
  &=& \frac{3}{5} \left(\frac{6\pi^2}{\gamma}\right)^{2/3}
    \left[\left(\rho_n^{5/3} + \rho_p^{5/3}\right)
      + \frac{5\pi^2}{3 \hbar^4}\left(\frac{\gamma}{6\pi^2}\right)^{4/3} T^2
     \left(\rho_n^{1/3} {m_n^*}^2 + \rho_p^{1/3} {m_p^*}^2\right)\right]_{x=0}
               \nonumber  \\
  & & - \frac{3}{5} \left(\frac{6\pi^2}{\gamma}\right)^{2/3}
     \left[\frac{5}{3} \rho_n^{5/3}
      + \frac{5\pi^2}{3 \hbar^4} \left(\frac{\gamma}{6\pi^2}\right)^{4/3}
     \rho_n^{1/3} {m_n^*}^2 \left(\frac{1}{3} - 2 b \rho_n \frac{m_n^*}{m_n}
        \right) T^2 \right]_{x=0}
                \nonumber  \\   & & \hspace{0.3in} \times
     \left[\left(\frac{e^y}{1 + e^y}\right) \left(\frac{R - R_n}{a}\right)
         + \left(\frac{3 + \pi^2 \left(\frac{a}{R}\right)^2}
                   {1 + \pi^2 \left(\frac{a}{R}\right)^2}\right)
              \left(\frac{R_n - R}{R}\right)\right]
               \nonumber  \\
  & & - \frac{3}{5} \left(\frac{6\pi^2}{\gamma}\right)^{2/3}
     \left[\frac{5}{3} \rho_p^{5/3}
      + \frac{5\pi^2}{3 \hbar^4} \left(\frac{\gamma}{6\pi^2}\right)^{4/3}
     \rho_p^{1/3} {m_p^*}^2 \left(\frac{1}{3} - 2 b \rho_p \frac{m_p^*}{m_p}
        \right) T^2 \right]_{x=0}
                \nonumber  \\   & & \hspace{0.3in} \times
     \left[\left(\frac{e^y}{1 + e^y}\right) \left(\frac{R - R_p}{a}\right)
         + \left(\frac{3 + \pi^2 \left(\frac{a}{R}\right)^2}
                   {1 + \pi^2 \left(\frac{a}{R}\right)^2}\right)
              \left(\frac{R_p - R}{R}\right)\right]
      + \cdots
\end{eqnarray}
with
\begin{eqnarray}
 \tau_q &=& \frac{3}{5} \left(\frac{6\pi^2}{\gamma}\right)^{2/3}
    \left[\rho_q^{5/3}
      + \frac{5\pi^2}{3 \hbar^4}\left(\frac{\gamma}{6\pi^2}\right)^{4/3} T^2
         \rho_q^{1/3} {m_q^*}^2
      + \cdots \right]          \nonumber  \\
  &=& \frac{3}{5} \left(\frac{6\pi^2}{\gamma}\right)^{2/3}
    \left[\rho_q^{5/3}
      + \frac{5\pi^2}{3 \hbar^4}\left(\frac{\gamma}{6\pi^2}\right)^{4/3} T^2
     \rho_q^{1/3} {m_q^*}^2 \right]_{t=0}
               \nonumber  \\
  & & - \frac{3}{5} \left(\frac{6\pi^2}{\gamma}\right)^{2/3}
     \left[\frac{5}{3} \rho_q^{5/3}
      + \frac{5\pi^2}{3 \hbar^4} \left(\frac{\gamma}{6\pi^2}\right)^{4/3}
     \rho_q^{1/3} {m_q^*}^2 \left(\frac{1}{3} - 2 b \rho_q \frac{m_q^*}{m_q}
      \right) T^2 \right]_{t=0}
                \nonumber  \\   & & \hspace{0.3in} \times
     \left[\left(\frac{e^y}{1 + e^y}\right) \left(\frac{R - R_q}{a}\right)
         + \left(\frac{3 + \pi^2 \left(\frac{a}{R}\right)^2}
                   {1 + \pi^2 \left(\frac{a}{R}\right)^2}\right)
              \left(\frac{R_q - R}{R}\right)\right]
      + \cdots                  \nonumber  \\
  &=& \frac{3}{5} \left(\frac{6\pi^2}{\gamma}\right)^{2/3}
    \left[\rho_q^{5/3}
        + \frac{5\pi^2}{3 \hbar^4}\left(\frac{\gamma}{6\pi^2}\right)^{4/3} T^2
        \rho_q^{1/3} {m_q^*}^2 \right]_{t=0}
                \nonumber  \\
  & & + \frac{3}{5} \left(\frac{6\pi^2}{\gamma}\right)^{2/3}
        \left[\frac{5}{3} \rho_q^{5/3}
        + \frac{5\pi^2}{3 \hbar^4} \left(\frac{\gamma}{6\pi^2}\right)^{4/3}
        \rho_q^{1/3} {m_q^*}^2 \left(\frac{1}{3} - 2 b \rho_q \frac{m_q^*}{m_q}
        \right) T^2 \right]_{t=0}
                \nonumber  \\   & & \hspace{0.3in} \times
        \left[\left(\frac{e^y}{1 + e^y}\right) \left(\frac{R_q - R}{a}\right)
         - \left(\frac{3 + \pi^2 \left(\frac{a}{R}\right)^2}
                   {1 + \pi^2 \left(\frac{a}{R}\right)^2}\right)
                \left(\frac{R_q - R}{R}\right)\right]
        + \cdots
\end{eqnarray}
and
\begin{eqnarray}
 \rho_q \tau_q
  &=& \rho_q \frac{3}{5} \left(\frac{6\pi^2}{\gamma}\right)^{2/3}
     \left[\rho_q^{5/3}
      + \frac{5\pi^2}{3 \hbar^4}\left(\frac{\gamma}{6\pi^2}\right)^{4/3} T^2
         \rho_q^{1/3} {m_q^*}^2 + \cdots \right]          \nonumber  \\
  &=& \rho_q \frac{3}{5} \left(\frac{6\pi^2}{\gamma}\right)^{2/3}
     \left[\rho_q^{5/3}
      + \frac{5\pi^2}{3 \hbar^4}\left(\frac{\gamma}{6\pi^2}\right)^{4/3} T^2
         \rho_q^{1/3} {m_q^*}^2 \right]_{x=0}
               \nonumber  \\
  & & - \rho_q \frac{3}{5} \left(\frac{6\pi^2}{\gamma}\right)^{2/3}
     \left[\frac{8}{3} \rho_q^{5/3}
      + \frac{5\pi^2}{3 \hbar^4} \left(\frac{\gamma}{6\pi^2}\right)^{4/3}
     \rho_q^{1/3} {m_q^*}^2 \left(\frac{4}{3} - 2 b \rho_q \frac{m_q^*}{m_q}
        \right) T^2 \right]_{x=0}
                \nonumber  \\   & & \hspace{0.3in} \times
     \left[\left(\frac{e^y}{1 + e^y}\right) \left(\frac{R - R_q}{a}\right)
         + \left(\frac{3 + \pi^2 \left(\frac{a}{R}\right)^2}
                   {1 + \pi^2 \left(\frac{a}{R}\right)^2}\right)
              \left(\frac{R_q - R}{R}\right)\right]
      + \cdots                   \nonumber  \\
  &=& \rho_q \frac{3}{5} \left(\frac{6\pi^2}{\gamma}\right)^{2/3}
     \left[\rho_q^{5/3}
      + \frac{5\pi^2}{3 \hbar^4}\left(\frac{\gamma}{6\pi^2}\right)^{4/3} T^2
         \rho_q^{1/3} {m_q^*}^2 \right]_{t=0}
               \nonumber  \\
  & & + \rho_q \frac{3}{5} \left(\frac{6\pi^2}{\gamma}\right)^{2/3}
     \left[\frac{5}{3} \rho_q^{5/3}
      + \frac{5\pi^2}{3 \hbar^4} \left(\frac{\gamma}{6\pi^2}\right)^{4/3}
     \rho_q^{1/3} {m_q^*}^2 \left(\frac{1}{3} - 2 b \rho_q \frac{m_q^*}{m_q}
        \right) T^2 \right]_{t=0}
                \nonumber  \\   & & \hspace{0.3in} \times
     \left[\left(\frac{e^y}{1 + e^y}\right) \left(\frac{R_q - R}{a}\right)
         - \left(\frac{3 + \pi^2 \left(\frac{a}{R}\right)^2}
                   {1 + \pi^2 \left(\frac{a}{R}\right)^2}\right)
              \left(\frac{R_q - R}{R}\right)\right]
               \nonumber  \\
  & & + \rho_q \frac{3}{5} \left(\frac{6\pi^2}{\gamma}\right)^{2/3}
     \left[\rho_q^{5/3}
      + \frac{5\pi^2}{3 \hbar^4} \left(\frac{\gamma}{6\pi^2}\right)^{4/3}
     \rho_q^{1/3} {m_q^*}^2 T^2 \right]_{t=0}
                \nonumber  \\   & & \hspace{0.3in} \times
     \left[\left(\frac{e^y}{1 + e^y}\right) \left(\frac{R_q - R}{a}\right)
         - \left(\frac{3 + \pi^2 \left(\frac{a}{R}\right)^2}
                   {1 + \pi^2 \left(\frac{a}{R}\right)^2}\right)
              \left(\frac{R_q - R}{R}\right)\right]
      + \cdots
\end{eqnarray}

In Weizacker mass formular it might be better expanding in terms
of $(R_n - R_p)/R$ rather than $(R_n - R_p)/a$ since the first one
is independent of $A$ while the second one is dependent on $A$.

\end{document}